\documentclass{article}

\usepackage{microtype}
\usepackage{graphicx}
\usepackage{subfigure}
\usepackage{booktabs} 
\usepackage{amsfonts}
\usepackage{hyperref}

\usepackage[accepted]{icml2021}

\icmltitlerunning{}

\newcommand{\physgan}{\textsc{PhysioGAN}\xspace}
\newcommand{\name}{\textsc{PhysioGAN}\xspace}
\newcommand{\nametitle}{PhysioGAN\xspace}

\usepackage{graphicx}
\usepackage{todonotes}
\usepackage{comment}
\usepackage[normalem]{ulem}

\usepackage{xcolor}
\usepackage{xspace}
\usepackage{booktabs}
\colorlet{kw}{blue}
\definecolor{com}{rgb}{0,0.6,0.3}

\usepackage{amsmath}
\usepackage{algorithm}
\usepackage{algorithmic}
\usepackage{graphicx}
\usepackage{bm}
\usepackage{caption}
\usepackage{arydshln}
\usepackage{enumitem}
\setitemize{noitemsep,topsep=0pt,parsep=0pt,partopsep=0pt}

\begin{document}

\twocolumn[
\icmltitle{\nametitle: Training High Fidelity Generative Model\\ for Physiological Sensor Readings}




\begin{icmlauthorlist}
\icmlauthor{Moustafa Alzantot}{to}
\icmlauthor{Luis Garcia}{to}
\icmlauthor{Mani Srivastava}{to}
\end{icmlauthorlist}

\icmlaffiliation{to}{University of California, Los Angeles}

\icmlcorrespondingauthor{Moustafa Alzantot}{malzantot@ucla.edu}
\icmlcorrespondingauthor{Luis Garcia}{gacialuis@ucla.edu}
\icmlcorrespondingauthor{Mani Srivastava}{mbs@ucla.edu}

\icmlkeywords{Generative Models, Neural Networks, Deep Learning, Physiological Sensors, Adversarial Training, Synthetic Datasets, Machine Learning}

\vskip 0.3in
]



\printAffiliationsAndNotice{}  

\begin{abstract}
Generative models such as the variational autoencoder (VAE) and the generative adversarial networks (GAN) have proven to be incredibly powerful for the generation of synthetic data that preserves statistical properties and utility of real-world datasets, especially in the context of image and natural language text.  Nevertheless, until now, there has no successful demonstration of how to apply either method for generating useful physiological sensory data. The state-of-the-art techniques in this context have achieved only limited success. We present \name, a generative model to produce high fidelity synthetic physiological sensor data readings. \name consists of an encoder, decoder, and a discriminator. We evaluate \name against the state-of-the-art techniques using two different real-world datasets:  ECG classification and activity recognition from motion sensors datasets. We compare \name to the baseline models not only the accuracy of class conditional generation but also the sample diversity and sample novelty of the synthetic datasets. We prove that \name generates samples with higher utility than other generative models by showing that classification models trained on only synthetic data generated by \name have only 10\% and 20\% decrease in their classification accuracy relative to classification models trained on the real data. Furthermore, we demonstrate the use of \name for sensor data imputation in creating plausible results.

\end{abstract}

\section{Introduction}
\label{sec:introduction}



Improved techniques for training generative models is a rapidly growing area of research. Over the past few years, the machine learning research community has made significant leaps forward towards this goal. This wave of success has been mainly driven by the advent of new training techniques such as the variational autoencoder (VAE)~\cite{kingma2013auto} and the generative adversarial networks (GAN)~\cite{goodfellow2014generative}. Through GANs and VAEs--as well as  their improved versions~\cite{salimans2016improved, brock2018large, razavi2019generating}--we are now capable of producing high fidelity, large-scale images with unprecedented levels of quality. GANs and VAEs have also been proven useful in a variety of applications such as generating photorealistic super-resolution images from  low-resolution images~\cite{ledig2017photo}, learning a disentangled latent space representation (which is valuable for content manipulation)~\cite{chen2016infogan, higgins2017beta}, and generating realistic images from text descriptions~\cite{reed2016generative}.

However, most of the generative model research has focused on training models for images ~\cite{radford2015unsupervised, salimans2016improved, brock2018large, razavi2019generating}  and, more recently, text datasets~\cite{ yu2017seqgan, juefei2018rankgan, wang2018sentigan, zhu2018texygen}. Only a few  of the existing works have studied, with limited success, how to learn a generative model for time series data such as sensor readings. Data from physiological sensors, e.g.,  electrocardiograms (ECGs) and fitness tracking sensors, are now prevalent in many applications for health monitoring and patient diagnosis.  A good generative model for physiological sensor readings is important for many high-utility applications in the medical domain. To name a few \textit{potential} applications, GANs have been successfully used to boost the performance of semi-supervised learning classification models that learn from a small set of labeled examples and a larger set of unlabeled examples on image datasets ~\cite{salimans2016improved}. In the medical domain, semi-supervised learning is highly desirable since labeling medical sensors readings--e.g., whether an ECG signal segment is normal or abnormal--can be costly and doable only by medical professionals. Further, GANs have been utilized to address concerns of privacy in the context of machine learning. Since machine learning models store information about training data, it has been shown that they can be reverse-engineered by an attacker~\cite{fredrikson2015model,shokri2017membership} to uncover sensitive information about the training data set. GANs have been used in combination with the differential privacy techniques~\cite{dwork2014algorithmic} to train accurate models with strong privacy guarantees against this kind of attacks~\cite{papernot2016semi}. Researchers have shown that GANs can produce synthetic datasets that can be used in place of the original real data while still being useful for performing analysis or even training newer models~\cite{uclanesl_dp_wgan, jordon2018pate}. These solutions are invaluable for researchers in the medical domain since the privacy-sensitive nature of medical datasets--along with their associated laws and regulations such as the `Institution Review Board` (IRB)--prevents researchers from sharing the data they collect~\cite{silberman2011burdens}. Unfortunately, the state-of-the-art methods for training generative models on sensor data readings are still far away from being able to satisfy the requirements of these applications.

The few efforts that have explored training generative models for sensor readings have focused on simple tasks with \textit{toy} datasets rather than meaningful, real-world tasks and datasets. For example, SenseGen ~\cite{alzantot2017sensegen} uses a recurrent neural network to train a generative model for accelerometer sensor readings using the maximum-likelihood objective. However, SenseGen was only capable of performing \textit{unconditional} generation and, thus, cannot control the attributes of the generator outputs. While the model can be easily extended to support conditional generation--as we will show in our experiments--we find that this training approach is not capable of delivering highly accurate conditional generation results.
We also studied the conditional variational recurrent autoencoder (CVRAE) approach for training a generative model for sensor data. The variational autoencoder maximizes an \textit{inexact} \textit{lower-bound} of the likelihood of training data is generally better at generating novel samples than traditional autoencoders~\cite{kingma2013auto}. Nevertheless, the accuracy of the conditional generation is also not high with this training approach. On the other hand, adversarially trained models using the GANs training framework are capable of producing more accurate conditionally generated samples. The RCGAN model~\cite{esteban2017real} has demonstrated how to use this approach to train a recurrent neural network generator for conditional generation of medical sensors data. However, as we show in our evaluation section, we find that despite having a very high accuracy with the conditional generation, the RCGAN model suffers from a lack of \textit{diversity}. We empirically show that the RCGAN model produces samples that are very similar and nearly identical within each class. The lack of sample diversity in GANs models is a well-known problem known as \textit{mode collapse}~\cite{theis2015note}--which is currently being addressed by the machine learning community~\cite{srivastava2017veegan}. Although mode collapse is not unique to generator models that produce sensor data, it is more severe in RCGAN because it utilizes a recurrent generator~\cite{unrolledgan}. Since the \textit{discriminator} used for RCGAN training does not provide any penalty when the generator produces repeated samples, the powerful recurrent generator tends to identify which subset of examples are good enough to fool the discriminator and simply repeats them--leading to a lack of sample diversity. Synthetic datasets that have samples suffering from either low generation accuracy or low diversity will have an equally poor performance when used as a replacement for real, \textit{private} data. 

A training approach is necessary that combines the merits of variational recurrent autoencoder approach with the GANs approach to produce a synthetic dataset that has both high conditional generation accuracy as well as a high diversity of samples. We introduce \name, an approach for training generative models that fulfills these objectives. 
\name consists of three different components: an \textit{encoder}, a \textit{decoder} and a \textit{discriminator}. 
Together, the \textit{encoder} and the \textit{decoder} form a conditional variational recurrent autoencoder (CVRAE) similar to the  CVRAE model--which we consider as a baseline. To improve the accuracy of conditional generation by the CVRAE, we introduce two additional training objectives provided by a \textit{discriminator}: the \textit{adversarial loss} and the \textit{feature matching} loss. The \textit{discriminator} itself is trained as a multi-class classifier that predicts the class label of real data and attempts to identify  ``fake" samples produced by the \textit{generator}. To address the issue of \textit{mode collapse}, we introduce an additional \textit{diversity loss}  that urges the \textit{generator} to maximize the mutual information between its output and the latent space noise used to generate them. Therefore, the \textit{diversity loss} penalizes the \textit{decoder}--which acts as a \textit{generator}--when it generates identical samples.We improve the training stability by using an annealing approach where the model training cost function \textit{softly} changes from a pure autoencoder loss to the new loss that combines the variational autoencoder loss with the \textit{feature matching loss}, \textit{diversity loss} and \textit{adversarial loss}.

We evaluate \name against four different baselines: the conditional-recurrent neural network generator (CRNN) (which is an extension of ~\cite{alzantot2017sensegen} that allows for conditional generation), the conditional variational recurrent autoencoder (CVRAE), the conditional recurrent GANs (RCGAN)~\cite{esteban2017real}, and a variation of RCGAN that has a modified \textit{auto-regressive} generator (RCGAN-AR). We conduct our experiments using two real-world tasks and datasets. The first dataset is the ``AFib classification dataset"~\cite{yildirim2018arrhythmia}, which is a dataset of ECG signal segments. Each segment is labeled as either ``Normal" or ``Atrial Fibrillation (AFib)", which is a major kind of irregular heartbeat (also known as arrhythmia) that can lead to heart failures and possibly death. The second dataset is a human activity recognition (HAR) dataset~\cite{anguita2013public} based on motion sensors such as the accelerometer and the gyroscope commonly found in wearable fitness tracking devices. The HAR dataset represents a dataset for multi-class classification with 6 different kinds of activities that can be grouped into two major groups. Because each group has 3 types of activities that are highly similar to each other and, learning how to conditionally generate samples is a difficult challenge. Further, the HAR dataset introduces the challenge of multi-channel data since each data sample has 6 different axes corresponding to correlated sensor readings. In addition to providing a visualization and qualitative comparison of samples produced by each model, we quantitatively evaluated the 5 models (\name and the 4 other baselines) based on their \textit{conditional generation accuracy}, the \textit{diversity} of generated samples, as well as the \textit{novelty} of samples to ensure that the model is not \textit{simply} reproducing the same samples as those observed during the training. In addition to those metrics, we use an additional metric to measure the \textit{utility}~\cite{esteban2017real, ravuri2019classification} of the synthetic dataset produced by each generator. The utility of a synthetic dataset measures how well the dataset can be used to train a classification model using \textit{only} the synthetic data by validating its performance against the accuracy of a model trained on the \textit{real dataset}. We demonstrate that classification models trained on \textit{only} synthetic data generated by \name have only 10\% and 20\% decrease in their classification accuracy than classification models trained on the real data while significantly outperforming existing approaches in terms of diversity and novelty. This significantly surpasses the accuracy of models trained on synthetic datasets generated by other baseline methods that relied only on the vanilla GAN or VAE for model training. Compared to other methods, \name attains a good balance between the accuracy of samples and their diversity.  We further evaluate the utility of \name in the context of sensor data imputation against state-of-the-art imputation techniques. Our results show that \name is capable of repairing corrupted time-series with missing the values with a higher degree of realism than other methods, both neural network-based and traditional methods for sensor data imputation.

\textbf{Contributions:} Our contributions are multi-fold:
\begin{itemize}[noitemsep,topsep=0pt,parsep=0pt,partopsep=0pt]
\item First, we identify common issues that state-of-the-art models currently suffer from by evaluating existing approaches and available baselines of generative models for time-series sensor readings. 
\item Second, we provide a novel model training method, \name, that combines both generative adversarial networks and variational recurrent autoencoders to train a generative model for sensor readings that produces samples with high accuracy and high diversity. 
\item Third, we evaluate \name and other baselines on two different datasets and show that \name is capable of producing a synthetic dataset that are self-sufficient for training models with moderate decrease in their accuracies than models trained on real (not synthetic) data.
\item Additionally, all of our model implementations and experiments are available as open-source~\footnote{\url{https://github.com/nesl/physiogan}} to promote further research in this important direction of research.
\end{itemize}

\noindent The rest of this paper is organized as follows: Section ~\ref{sec:related} provides a background on the different kinds of generative models such as GANs and VAEs and also summarizes the related work in training generative models for time-series sensor data readings. Section ~\ref{sec:methodology} describes our model architecture and training procedure details. Section ~\ref{sec:evaluation} includes the results of our evaluation experiments. Finally, Section~\ref{sec:discussion} concludes the paper.

\section{Background and Related Work}
\label{sec:related}
In this section, we present the preliminary information necessary to understand the \name model as well as the works directly related to the scope of this paper. 
\subsection{Background}
We first present an overview of the two state-of-the-art frameworks for training generative models: generative adversarial networks (GANs), and variational autoencoders (VAEs).
\subsubsection*{Generative Adversarial Networks}
Generative adversarial networks (GANs)~\cite{goodfellow2014generative} were presented as  a framework for training generative models. GANs simultaneously train two separate models through an adversarial game. The first model--called the \textit{generator}, $\mathbf{G}$--learns the distribution of training data. Instead of producing an explicit probability density value, the goal of the \textit{generator} is to directly produce samples from the distribution it learned. The input to  the \textit{generator} $\mathbf{G}$ is a noise vector, $\mathbf{z}$, sampled from an arbitrary chosen prior noise distribution $p_{\mathbf{z}}(\mathbf{z})$, i.e.,  $\mathbf{z} \sim p_{\mathbf{z}}(\mathbf{z})$. The noise distribution $p_{\mathbf{z}}(\mathbf{z})$ is typically chosen as the standard Gaussian distribution $\mathcal{N}(0, \mathbf{I})$. The \textit{generator} function, $\mathbf{G}(\mathbf{z})$, translates that random noise into fake samples that match the real samples drawn from the training dataset. The second model is referred to as the \textit{discriminator}, $\mathbf{D}$. The \textit{discriminator} distinguishes between the fake samples produced by the \textit{generator} and the real samples from the training dataset. $D(\mathbf{x})$ represents the probability that the input $\mathbf{x}$ is drawn from the real data distribution rather than coming from the generator outputs. The training objective of the \textit{discriminator}, $\mathbf{D}$, is to increase its accuracy in distinguishing between those two sets of samples. On the other hand, the training objective of \textit{generator}, $\mathbf{G}$, is to fool the \textit{discriminator} by producing fake samples that look sufficiently realistic such that it becomes harder for the \textit{discriminator} to identify them. This training procedure can be mathematically formalized as $\mathbf{D}$ and $\mathbf{G}$ playing a two-player min-max game with the following value function $V(\mathbf{G}, \mathbf{D})$:

\begin{equation}
\begin{split}
\label{eqn:gan}
\min_G \max_D V(G, D) = & \mathbb{E}_{\mathbf{x} \sim p_{\text{data}}(\mathbf{x})}[\log D(\mathbf{x})] + \\ & \mathbb{E}_{\mathbf{z} \sim p_{\mathbf{z}}(\mathbf{z})}[\log (1 - D(G(\mathbf{z})))]
\end{split}
\end{equation}

Conditional GANs~\cite{mirza2014conditional} extend the original GANs~\cite{goodfellow2014generative} models to generate samples that are conditioned on a given class label attribute $\mathbf{y}$. This can be achieved by feeding $\mathbf{y}$ as additional input to both the \textit{discriminator} $\mathbf{D}$ and \textit{generator} $\mathbf{G}$. Therefore, the objective function to the two-player min-max game becomes:
\begin{equation}
\begin{split}
\label{eqn:cgan}
\min_G \max_D  V(G, D) =   & \mathbb{E}_{\mathbf{x},\mathbf{y} \sim p_{\text{data}}(\mathbf{x} | \mathbf{y})}[\log D(\mathbf{x} | \mathbf{y})] + \\
&  \mathbb{E}_{\substack{\mathbf{z} \sim p_{\mathbf{z}}(\mathbf{z}) \\ \mathbf{y} \sim Cat(\{1,..,L\})}}[\log (1 - D(G(\mathbf{z} | \mathbf{y})))]
\end{split}
\end{equation}

Over the past few years, many extensions to GANs have been proposed~\cite{radford2015unsupervised, salimans2016improved, arjovsky2017wasserstein}, and they have been successfully applied in a variety of domains such as generating realistic super-resolution images~\cite{ledig2017photo}, image in-painting~\cite{demir2018patch}, and image synthesis based on text description~\cite{reed2016generative, xu2018attngan}. 
However, despite the recent success of GANs, 
successful training of GANs remains a challenge as it requires finding the Nash equilibrium between two non-cooperating players $\mathbf{G}$ and $\mathbf{D}$.  The Nash equilibrium happens when the cost of each player is minimized with respect to its own parameters. However, since GANs training is done by applying gradient descent to alternately minimize both the discriminator loss and the generator loss, there is no guarantee that this training approach will converge as minimizing one of the losses may increase the other. Therefore, researchers have suggested various tricks to improve the stability of GANs training such as architecture guidelines for both generators and discriminators~\cite{radford2015unsupervised}, mini-batch discrimination~\cite{salimans2016improved}, and historical averaging of model weights~\cite{salimans2016improved}. Wasserstein GAN~\cite{arjovsky2017wasserstein, gulrajani2017improved} is a recent improvement that replaces the \textit{discriminator} by a critic and uses either weight clipping~\cite{arjovsky2017wasserstein} or gradient-penalties~\cite{gulrajani2017improved} to enforce a Lipschitz constraint to improve the training stability.


\subsubsection*{Variational AutoEncoders}
In addition to GANs, the variational autoencoder (VAE)~\cite{kingma2013auto} is another state-of-the-art framework for training generative models. Unlike GANs, where the generator, $\mathbf{G}$, is trained to fool the discriminator, the objective of VAE training is based on maximum likelihood estimation. Intuitively, increasing the likelihood of the generator model to produce training data samples will also increase its capability of generating samples that are similar to the training data.

A major assumption VAEs make is that the data points $\mathbf{x}$  are generated in response to some latent code variable $\mathbf{z}$ that are drawn from a prior distribution $p_{\mathbf{z}}(\mathbf{z})$. According to the law of total probability, the likelihood of one example $\mathbf{x}^{(i)}$ can be expressed as:
\begin{equation}
     p_{\theta}(\mathbf{x}^{(i)}) = \int{p_{\theta}(\mathbf{x}^{(i)}|\mathbf{z}) \, p_{\mathbf{z}}(\mathbf{z}) \, dz}
     \label{eqn:ll}
\end{equation}
\noindent where the model function $p_{\theta}(\mathbf{x}^{(i)}|\mathbf{z})$ acts as a \textit{decoder} that produces the likelihood of sample $\mathbf{x}^{(i)}$  to be generated according to the latent space value of $\mathbf{z}$. When the decoder is implemented as a neural network, which is capable of being a universal function approximator, it can translate the arbitrarily chosen distribution of the latent space variables into the learned data distribution.  The prior distribution of the noise $p_{\mathbf{z}}(\mathbf{z})$ is typically chosen as the standard Gaussian distribution $\mathcal{N}(\mathbf{0}, \mathbf{I})$.  However, this likelihood integral in equation ~\ref{eqn:ll} is intractable because there are many possible values of $\mathbf{z}$ and most of them will not have a significant likelihood of producing the given example $\mathbf{x}^{(i)}$. VAEs address this issue by introducing another network called the \textit{inference} or \textit{recognition} model, $q_{\phi}(\mathbf{z}|\mathbf{x})$, that approximates the true posterior $p_{\theta}(\mathbf{z}|\mathbf{x})$. VAEs use the expected log-likelihood of training samples under this approximate posterior: 
\begin{equation}
\label{eqn:vaell}
\begin{split}
\log\;\bigl(p_{\theta}(\mathbf{x}^{(i)})\bigr) = & \;\mathbb{E}_{\mathbf{z} \sim q_{\phi}(\mathbf{z} | \mathbf{x}^{(i)})} \bigl[\log \;p_{\theta}({\mathbf{x}^{(i)}}) \bigr]\\
= &\; \mathbb{E}_{\mathbf{z} \sim q_{\phi}(\mathbf{z} | \mathbf{x}^{(i)})} \bigl[ {\log \, \frac{p_{\theta}(\mathbf{x}^{(i)}|\mathbf{z}) p(\mathbf{z})}{p(\mathbf{z} | \mathbf{x}^{(i)})}} \bigr]\\
= &\;  \mathbb{E}_{\mathbf{z} \sim q_{\phi}(\mathbf{z} | \mathbf{x}^{(i)})} \left[ {\log \, \frac{p_{\theta}(\mathbf{x}^{(i)}|\mathbf{z}) p(\mathbf{z})}{p(\mathbf{z} | \mathbf{x}^{(i)})}} 
\frac{q_{\phi}(\mathbf{z}|\mathbf{x}^{(i)})}{q_{\phi}(\mathbf{z}|\mathbf{x}^{(i)})}\right]
\end{split}
\end{equation}
This equation can be simplified as:
\begin{multline}
\label{eqn:vae2}
\log\bigl(p_{\theta}(\mathbf{x}^{(i)})\bigr) =   \mathbb{E}_{\mathbf{z} \sim q_{\phi}(\mathbf{z} | \mathbf{x}^{(i)})} \big[  \log \, p_{\theta}(\mathbf{x}^{(i)} | \mathbf{z})) \quad-  \\
 D_{KL}\big( q_{\phi}(\mathbf{z} | \mathbf{x}^{(i)}) || p_{\mathbf{z}}(\mathbf{z}) \big] \\
 + D_{KL} \bigl(q_{\phi}(\mathbf{z} | \mathbf{x}^{(i)}) || p_{\theta}(\mathbf{z}| \mathbf{x}^{(i)}) \bigr)  
\end{multline}

\noindent where $D_{KL} $ is the KL-divergence function that measures the distance between two probability distributions, i.e.,
\[ D_{KL}(p || q) = \int{p(x) \, \log {( \frac{p(x)}{q(x)})} \, dx}  \]
\noindent This allows for equation \ref{eqn:vae2} to be rewritten as:
\begin{equation}
\begin{split}
\label{eqn:vae3}
\log \bigl(p_{\theta}(\mathbf{x}^{(i)})\bigr) = &\; \mathcal{L}_{ELBO}\bigl(\mathbf{x}^{(i)}; \theta, \phi \bigr)  + \\  & D_{KL} \big(q_{\phi}(\mathbf{z} | \mathbf{x}^{(i)}) || p_{\theta}(\mathbf{z}| \mathbf{x}^{(i)}) \big) 
\end{split}
\end{equation}
Since The KL-divergence term is always non-negative, i.e. $D_{KL}(p, q) \geq 0$, the first right hand side term--called the \textit{evidence lower bound} ($\mathcal{L}_{ELBO}$),  will constitute a lower-bound for the log-likelihood function, i.e.,  
\[ \mathcal{L}_{ELBO}\bigl(\mathbf{x}^{(i)}; \theta, \phi\bigr) \leq \log p_{\theta}(\mathbf{x}) \]
VAE training maximizes this lower-bound term $\mathcal{L}_{ELBO}$. The difference between the lower bound $\mathcal{L}_{ELBO}$ and the true log-likelihood indicates the error due to replacing the exact intractable posterior $p_{\theta}(\mathbf{z}|\mathbf{x})$ by an the approximate posterior $q_{\phi}(\mathbf{z}|x) $ which is tractable to use with help of the `\textit{recognition}' network.
The training loss of the \textit{variational} autoencoder is equal to the negative of the \textit{evidence lower bound}. Therefore, the training loss is defined as:
\begin{equation}
\label{eqn:elbo}
    \begin{split}
\mathcal{L}_{vae}(\mathbf{x}^{(i)}; \phi, \theta) & = 
- \mathcal{L}_{ELBO}(\mathbf{x}^{(i)}; \theta, \phi) \\
& = \Big[ \overbrace{-\mathbb{E}_{\mathbf{z} \sim q_{\phi}(\mathbf{z} | \mathbf{x}^{(i)})} \bigl(\log \, p_{\theta}(\mathbf{x}^{(i)} | \mathbf{z})\bigr)}^{\text{Reconstruction loss}} \\
 & \; \; + \overbrace{D_{KL}\bigl( q_{\phi}(\mathbf{z} | \mathbf{x}^{(i)}) || p_{\mathbf{z}}(\mathbf{z})\bigr)}^{\text{Posterior loss}}  \Big]
\end{split}
\end{equation}

The  first part of the right-hand side in equation ~\ref{eqn:elbo}  represents the log-likelihood of the training sample $\mathbf{x}^{(i)}$ generated by the \textit{decoder} network from a latent space input vector $\mathbf{z}$ sampled from the \textit{recognition} network. This term represents the \textit{reconstruction error} of the training example after being fed through the encoder-decoder networks. The second term of the right-hand side represents the KL-divergence between the distribution of the latent space values produced by the \textit{recognition} network and a chosen prior distribution of the latent space values. Therefore, the KL-divergence term acts as a regularizer that encourages the \textit{encoder} to produces latent space values that match the given prior. Typically, the latent space prior $p_{\mathbf{z}}(\mathbf{z})$ is chosen as an isotropic Gaussian.

Calculating the gradients of the $\mathcal{L}_{ELBO}$ requires backpropagation through the stochastic sampling of the \textit{encoder} output. Therefore ~\cite{kingma2013auto} introduced the technique known as the `\textit{reparamterization trick}' where the approximate posterior sampling $\mathbf{z} \sim q_{\phi}\big(p_{\mathbf{z}}(\mathbf{z})|\mathbf{x}^{(i)}\big)$ is replaced with a differentiable transformation. The \textit{recognition} network produces the mean, $\mu$, and the standard deviation, $\sigma$, of the output distribution. The stochastic sampling of $\mathbf{z}$ can now be approximated using the following equation:
\begin{equation}
\begin{split}
    \mathbf{z} \sim q_{\phi}\big(p_{\mathbf{z}}(\mathbf{z})|\mathbf{x}^{(i)}\big) \approx \mu + \sigma \odot \epsilon
\end{split}
\end{equation}
\noindent where $\epsilon \sim \mathcal{N}(0, \mathbf{I})$ is an auxiliary noise variable sampled from the standard Gaussian distribution and $\odot$ is the hadmard, element wise, product operator.

\subsection{Related Work}
Generative models for data synthesis has been an active topic of research over the past few years. However, most of the research focus has been on generating high fidelity images~\cite{brock2018large,karras2019style, zhang2018self} and tabular datasets~\cite{uclanesl_dp_wgan,park2018data,xu2018synthesizing,choi2017generating,yahi2017generative}. For time-series datasets, the focus has been on the generation of natural language text~\cite{mikolov2010recurrent,yu2017seqgan,juefei2018rankgan,hu2017toward,wang2018sentigan} and music datasets~\cite{roberts2018hierarchical}. Much less effort has been placed towards the generation of physiological sensor readings. In the rest of this section, we briefly discuss the related work on the generation of time-series and sensor readings.

\subsubsection*{Generative Models for Time-series Generation}
The ability to generate high-quality human language text is essential for a variety of tasks such as machine translation and AI chatbots. RNNLM ~\cite{mikolov2010recurrent} uses the maximum likelihood estimate (MLE) to train a recurrent neural network to predict the next word given the previous word. However, MLE is not a perfect training objective due to the \textit{exposure bias} problem~\cite{huszar2015not} that leads to performance degradation at the generation time due to the discrepancy between the model inputs at training and inference. Scheduled sampling technique ~\cite{bengio2015scheduled} was proposed to increase generation quality. However, it has been found that it will have the negative effect of decreasing sample diversity--leading to another issue known as \textit{mode collapse}~\cite{hu2017toward, theis2015note} where the model generates samples that are too similar to each other.
 
 SeqGAN~\cite{yu2017seqgan} described an approach for training text generative models by modeling the \textit{generator} as a stochastic policy agent of reinforcement learning (RL) which is trained using a policy gradient algorithm~\cite{sutton2000policy}. TextGAN~\cite{zhang2016generating} uses a GANs framework to simultaneously train a recurrent neural network (RNN)-based \textit{generator} with a convolutional neural network (CNN)-based \textit{discriminator}. Instead of using the standard GANs training objective for a generator, TextGAN~\cite{zhang2016generating} uses feature matching~\cite{salimans2016improved} that matches the mean and variances of discriminator feature vectors between the real and synthetic sentences. A generative model that combines both a variational autoencoder and a discriminator based adversarial training was introduced in ~\cite{hu2017toward} to generate plausible text sequences whose attributes are controlled by learning a disentangled latent space representation. Our model design is inspired by this work due to the high quality of results it offered in controlled text generation. A more comprehensive literature review on using generative models for text can be found in ~\cite{lu2018neural, zhu2018texygen}.

To generate time-series data outside the domain of text and music, WaveNet~\cite{oord2016wavenet} uses a similar maximum-likelihood based objective to generate speech signals and SktechRNN~\cite{ha2017neural} uses a recurrent neural network model to produce sketch-based drawings of common objects. Each drawing is represented as time-series of paint-brush strokes. 

\subsubsection*{Generative Models for Sensor Readings Generation}
In the following section, we provide an overview of previous work in the domain of generating synthetic sensor data readings. 

\noindent\textbf{Maximum Likelihood-based Models.} SenseGen~\cite{alzantot2017sensegen} uses the maximum-likelihood objective to train a recurrent neural network with mixture density distribution (MDN) outputs to generate synthetic sensor readings. Their framework is trained and evaluated using accelerometer sensor dataset. However, SenseGen is not capable of performing \textit{conditional} generation as it does not provide a mechanism to control the attributes of the generator results. Besides, models trained with only maximum-likelihood objective exhibit exposure bias~\cite{huszar2015not} that reduce the quality of the generated signal because the model is trained to predict the next step of sequence without being encouraged to model the holistic features of the signal. SenseGen~\cite{alzantot2017sensegen} did not have an evaluation for the quality or utility of generator results and used only visual quality to demonstrate the success of their model. To the best of our knowledge, there is no prior work in using VAEs to train generative models for sensor readings. VAEs have the nice capability of doing \textit{inference} by \textit{encoding} the input sample into a distributed latent space vector that captures global features of the signal. The ability to do \textit{inference} jointly with \textit{generation} not only improves the quality and diversity of generator results but also can be handy in applications such as sensor data imputation to fill missing segments of input examples.

\noindent\textbf{Adversarially trained models.} SensoryGAN~\cite{wang2018sensorygans} uses the GANs adversarial training objective to train generative models for three kinds of human activities: staying, walking, and jogging. However, their technique requires training a separate generator with a different model architecture for each kind of activity rather than using a \textit{single generator} with a conditioning input. This makes their approach less generalizable for other tasks and datasets. Also, in their experiments, they train models to generate only three human activities (\textit{staying}, \textit{walking}, \textit{jogging}) which are considered \textit{coarsely grained} and strongly dissimilar from each other due to the significant degree of difference in motion intensity. A more solid experiment would be to include activities that are considered similar with only \textit{fine grained} differences such as \textit{walking, walking upstairs, and walking downstairs}, or \textit{normal} vs \textit{abnormal} ECG signals we are doing in our experiments. The RCGAN~\cite{esteban2017real} uses a recurrent discriminator to train another recurrent neural network to act as a conditional generator. Experiments of ~\cite{esteban2017real} were only done using small `toy` datasets and short sequences of ICU samples from low-frequency sensors, unlike the raw motion and ECG sensors we consider in our experiment. Furthermore, as we show later in our analysis, RCGAN suffers from two major disadvantages. First, it suffers heavily from \textit{`mode collapse`}, which means it produces synthetic datasets of samples with very low \textit{diversity} and are very similar to each other. Second, the generator model in RCGAN is not able to produce synthetic segments longer than the length it has observed during the training. This is because RCGAN does not use auto-regressive feedback during its generation. The recent work of ~\cite{smith2020conditional} cascades two generators to generate time-series data where the first generator generates RGB spectrogram 2D images while the second generator network translates the 2D image into time-series data. Similarly, ~\cite{zhang2020deep} uses an adversarially trained neural network to translate an input video into accelerometer time-series data. Our work is novel in its architecture and training objective.  Our model generates time-series data conditioned only on the classification label without requiring an input image or video to translate into time-series data. Unlike prior and parallel work, We also conduct a systemic evaluation of the synthetic data using different metrics (accuracy, novelty, diversity, utility).

Compared to prior work, our work is the first of its kind that combines variational autoencoder with adversarial networks for the task of \textit{conditional generation} of synthetic sensor readings. 
By using a novel training objective that combines variational autoencoder and generative adversarial networks, our model is able to generate high-quality synthetic datasets that are both accurate and diverse. 

\begin{figure*}[!th]
\centering
    \begin{minipage}{0.48\linewidth}
\includegraphics[width=\linewidth]{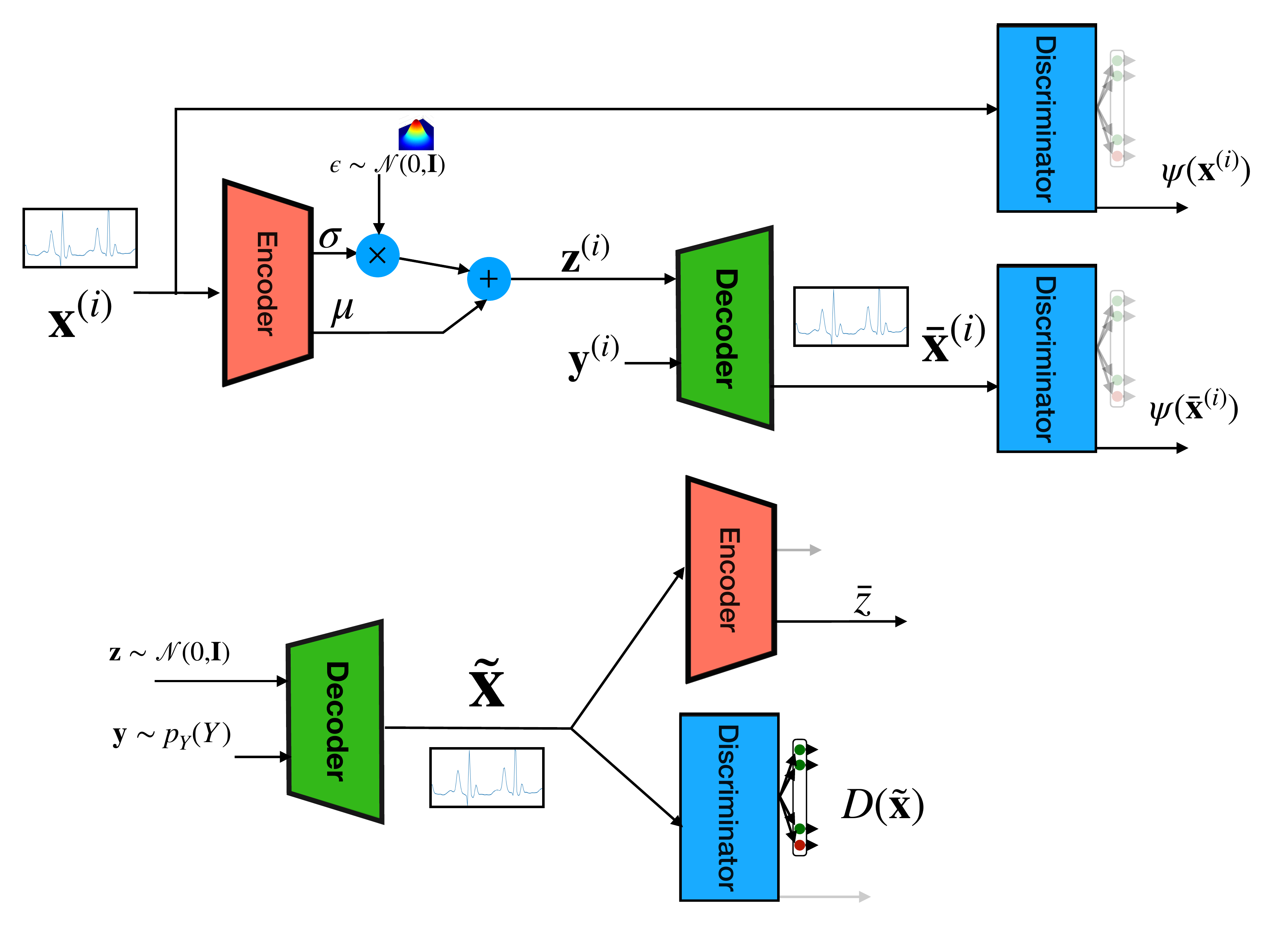}
\caption{Overview of model design for \name. \name consist of an \textit{encoder}, \textit{decoder} and \textit{discriminator} components. The different instances of the same component in this diagram are all sharing the same set of weights.}
\label{fig:overview-arch}
\end{minipage}
\hfill
    \begin{minipage}{0.48\linewidth}
\includegraphics[width=\linewidth]{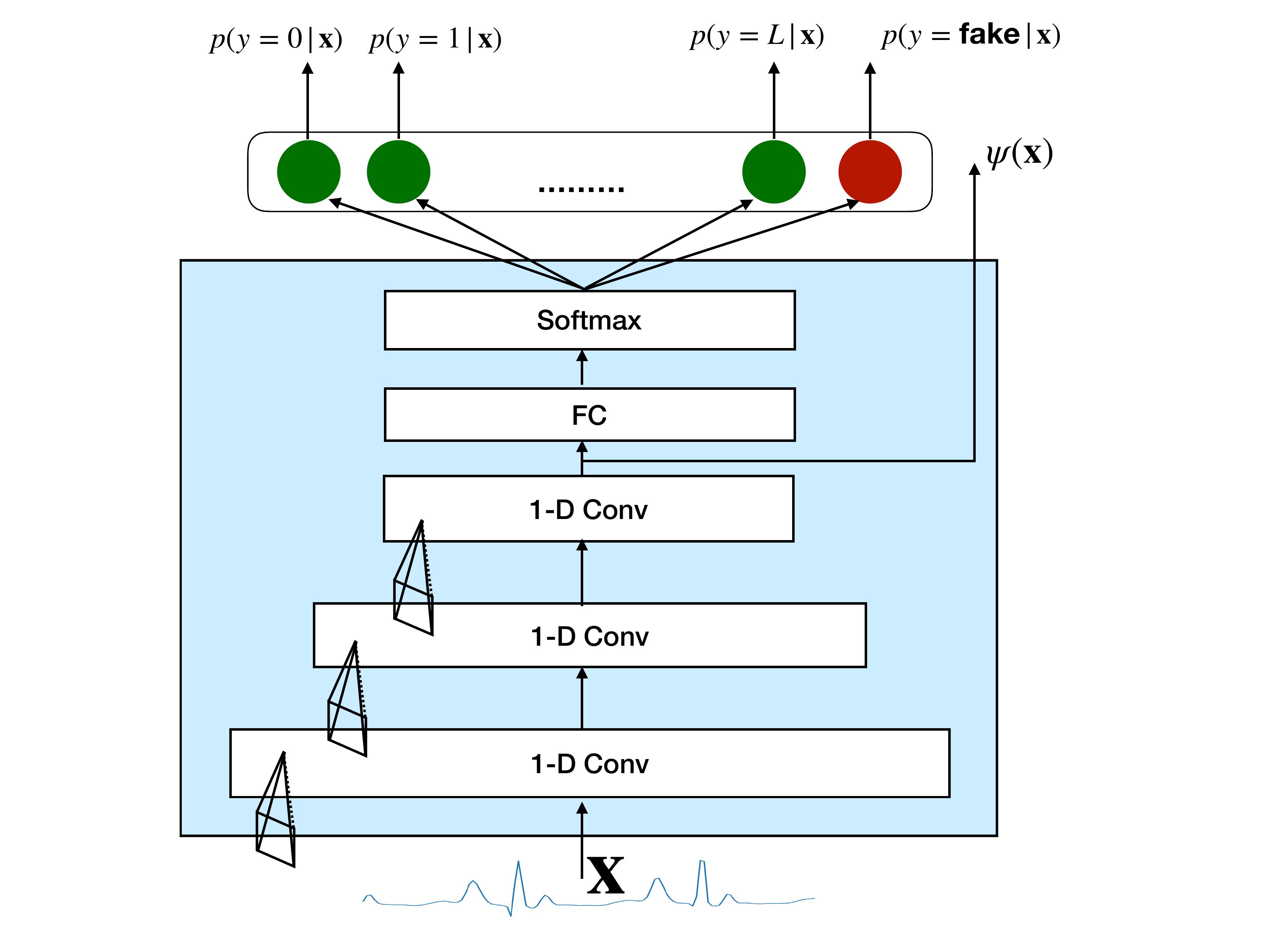}
\caption{Architecture of the \textit{Discriminator} model. The auxiliary output $\psi(\mathbf{x})$ is used to compute the \textit{feature matching} loss.}
\label{fig:discriminator}
\end{minipage}
\end{figure*}

\section{\name Methodology}
\label{sec:methodology}
In the following section, we describe our model architecture and training algorithm. 

\subsection{Objective}
\label{sec:objective}
Given a dataset of $N$ labeled time-series sensor data readings, $\mathcal{D}=\{(\mathbf{x}^{(i)}, y^{(i)})\}_{i=1}^{N}$,
Our goal is to build a `generator model' $\mathbf{G}$ that is capable of  synthetic real-valued time-series sensor readings \textit{conditioned on their class labels}. The sensor readings may be multi-dimensional at each time step. For example, each time-step may consist of multiple values measured across the different channels of the same sensor or multiple sensors sampled in time-synchronized intervals. Ideally, the synthetic data produced by the generator should look realistic and hard to distinguish from the real data sampled from the training set. They should also mimic the same distinguishing features and dynamics as the real data. Therefore, any analytic function computed over the synthetic data should return a value close to the returned value from the same function when computed over the real data. For instance, a machine learning classification model trained on a \textit{synthetic} dataset produced by our generator should yield good accuracy when tested using samples from the \textit{real} dataset.

\subsection{Notation}
Formally, the \textit{generator} function can be defined as:
\begin{equation}
\label{eqn:gen_def}
\begin{split}
    \tilde{\mathbf{x}} =& \;\mathbf{G}(\mathbf{z}, y; \theta) \\
    \mathbf{G}(\mathbf{z}, y; \theta):&\; \mathbb{R}^{N_z} \times  \mathcal{Y}  \rightarrow \{ \mathbb{R}^{T \times N_d} \}
\end{split}
\end{equation}

\noindent where $\mathbf{z} \in \mathcal{R}^{N_z}$  is an input random noise vector sampled from an arbitrarily chosen prior distribution (e.g., standard Gaussian) used as a source of variation to the deterministic \textit{generator} model function, and $y \in \mathcal{Y}$ is class label condition code used to specify the label of the samples we want as the generator output. For example, $\mathcal{Y}$ can be defined as
\begin{equation*}
 y \in \mathcal{Y} = \{\text{sitting}, \text{walking}, \text{running}\},
 \end{equation*}
\noindent to represent the activity label in a human activity classification dataset with those 3 classes of activities. $\theta$ represents the set of parameters of the generative models.
Each example in the dataset  $\mathbf{x}$ is a time-series with $T$ time-steps, i.e., 
\[ \mathbf{x} \in \mathcal{R}^{T \times N_d} \]
If we use the subscripted notation $\mathbf{x}_t$ to represent the time-series value at the single time step $t$, we can write $\mathbf{x}$ as
\[\mathbf{x} = \{\mathbf{x}_1,\mathbf{x}_2, \ldots, \mathbf{x}_T  \} \]
At each time step, $\mathbf{x}_t$ has $N_d$ real-valued numbers representing the values across the different channels of sensor readings (or the values across different time-synchronized sensors). For example, given a 3-axial accelerometer motion sensor, $N_d$ will be equal to three--corresponding to the three $X, Y, Z$ axes of the sensor readings. Thus, any $\mathbf{x_t}$ in this context can be represented as  
\[ {\mathbf{x}}_t = (\mathbf{x}_{t,1}, \mathbf{x}_{t,2},..., \mathbf{x}_{t,d})\in \mathcal{R}^{N_d} \quad \forall t \in [1,2,..T] \]
To summarize, equation \ref{eqn:gen_def} indicates that the generator learns how to translate an input noise vector and a condition class label 
into a time-series of real-valued numbers that looks realistic with respect to real sensor reading samples that match the designated condition label. Given this notation, we can now describe the model structure of \name.
\begin{figure*}[!th]
    \begin{minipage}{0.48\linewidth}
\includegraphics[width=\linewidth]{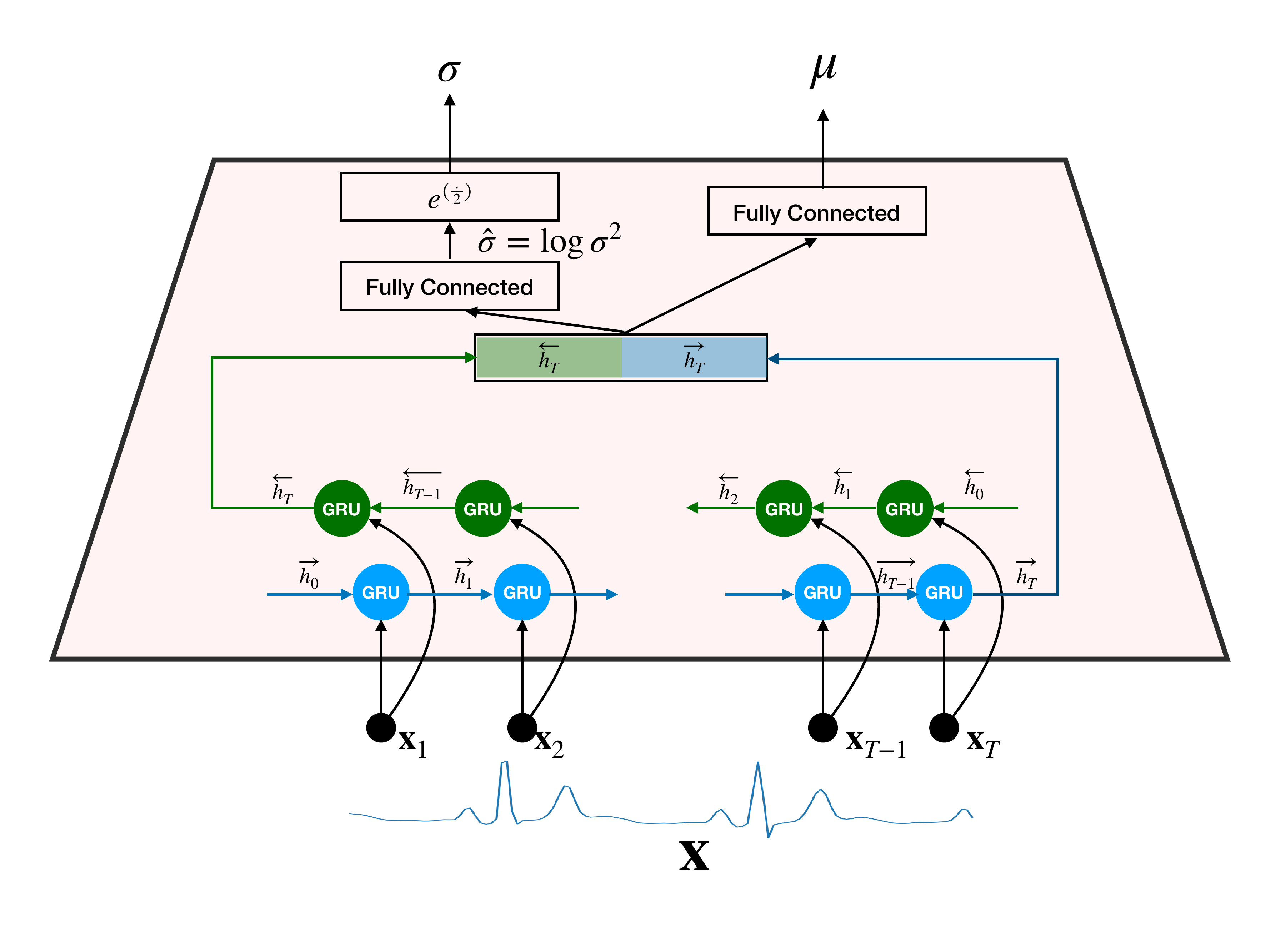}
\caption{Architecture of the \textit{Encoder} model.}
\label{fig:encoder}
\end{minipage}
    \begin{minipage}{0.48\linewidth}
\includegraphics[width=\linewidth]{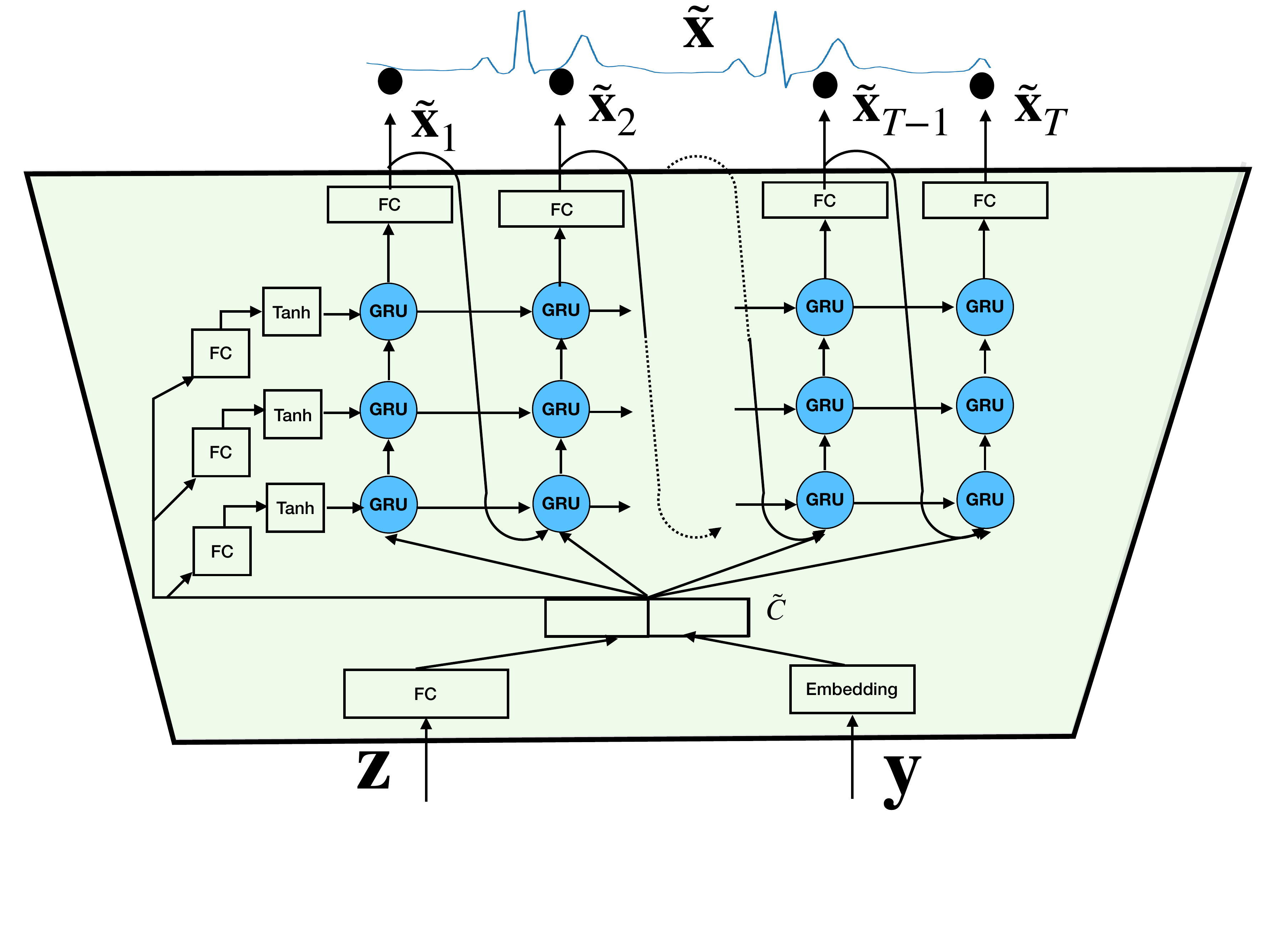}
\caption{Architecture of the \textit{Decoder} model.}
\label{fig:decoder}
\end{minipage}
\end{figure*}
\subsection{\name Model Structure}
\label{sec:model_structure}

 Our model takes advantage and draws inspirations from the recent success Generative adversarial networks (GAN) and variational autoencoders (VAE) have had for the generation of realistic time-series values in text and sketch-drawing domains~\cite{bowman2015generating, ha2017neural, hu2017toward}.  \name consists of three major components: an \textit{encoder}, a \textit{decoder}, and a \textit{discriminator}. Figure~\ref{fig:overview-arch} depicts the overview of \name's model structure and the interaction between the different components. Together, the \textit{encoder} and the \textit{decoder} form a sequence-to-sequence autoencoder that learns to take an input example $\mathbf{x}$ and reconstructs it back into $\bar{\mathbf{x}}$ after it has been mapped into a fixed-length  latent space vector $\mathbf{z}$. Both the \textit{encoder} and the \textit{decoder} are implemented as recurrent neural networks (RNNs). We utilize the VAE training objective--which maximizes the evidence lower bound $\mathcal{L}_{ELBO}$ in equation ~\ref{eqn:elbo}--since VAEs are better at generating new samples than traditional autoencoders~\cite{kingma2013auto}. Conditional generation from VAEs can be achieved by adding the attribute labels value $\mathbf{y}$ as an additional input to the \textit{decoder}. However, the element-wise \textit{reconstruction error} objective of the VAE does not encourage the \textit{decoder} to maintain the holistic features of the output time-series that are unique for each class label. Therefore, to improve the quality of conditional generation, we pair the VAE \textit{decoder} with a \textit{discriminator} model. We extend the VAE \textit{decoder} training objective--which acts as a \textit{generator}--in two ways. First, we use the GANs training approach to train the \textit{decoder} how to fool the \textit{discriminator} into accepting the samples it generates as \textit{realistic} and by using a multi-class classifier as our \textit{discriminator} it also penalizes the generation of the samples that looks realistic but belong to a wrong class. Second, we extend the \textit{reconstruction error} component of the training objective to include \textit{feature matching} between the features computed by the \textit{discriminator} on both the original input and reconstructed output of the discriminator. Feature matching encourages the decoder to not only preserve the element-wise similarity between autoencoder inputs and outputs but also to preserve the values of the holistic high-level  features that are specific for each class label. While introducing the GAN objective using the \textit{discriminator} into the \textit{decoder} training helps to improve the quality of generated samples. GANs are known to suffer from the issue of `\textit{mode collapse}`~\cite{theis2015note} where the \textit{generator} identifies which samples were able to fool the \textit{discriminator} successfully and  generates repeated copies of them.  In such a case, the model will exhibit  \textit{low diversity}, i.e., the model will generate samples that are too similar to each other.  To alleviate this issue and improve the diversity of generated samples, we feed the synthetic samples produced by the decoder from a noise vector $\mathbf{z}$ back into the \textit{encoder}, which reconstructs the noise vector $\bar{\mathbf{z}}$. Then, as an additional objective of the \textit{decoder} training, we introduce an additional loss term which measures the reconstruction error between $\bar{\mathbf{z}}$ and $\mathbf{z}$. This term encourages the \textit{decoder} to produce samples that are unique for each $\mathbf{z}$ value so that the \textit{encoder} will be able to approximately recover the value of $\mathbf{z}$ as $\bar{\mathbf{z}}$.

In the rest of this section, we describe the details of building individual components of \name as well as how to train them.

\subsubsection*{Encoder Design}
The \textit{encoder} translates the input time series sequence into a latent space vector sampled from a posterior distribution, i.e., $ \mathbf{z} \sim q_{\phi}(\mathbf{z} | \mathbf{x}) $. As shown in Figure ~\ref{fig:encoder}, our \textit{encoder} model is implemented using a bidirectional recurrent neural network that accepts an input time-series sequence $\mathbf{x} \in \mathbb{R}^{T \times N_d}$  and produces a latent vector $\mathbf{z} \in \mathbb{R}^{N_z}$. The bidirectional recurrent neural network consists of two recurrent neural networks that process the input sequence in the forward and backward direction, respectively. Each one of those recurrent neural networks evolves a hidden state vector $\overrightarrow{h}$, $\overleftarrow{h}$ while processing the input sequence time-step by time-step. We use the Gated Recurrent Unit (GRU)~\cite{cho2014properties,chung2015gated} implementation of the recurrent unit in our \textit{encoder}.
\begin{equation}
\begin{split}
\overrightarrow{o}_t, \overrightarrow{h}_{t} = GRU(\mathbf{x}_{t}, \overrightarrow{h}_{t-1}) \\
\overleftarrow{o}_t, \overleftarrow{h}_{t} = GRU(\mathbf{x}^{(reversed)}_{t}, \overleftarrow{h}_{t-1}) \\
\end{split}
\end{equation} 
where $\mathbf{x}_{t}^{(reversed)}$ is the input time-series $\mathbf{x}$ after being reversed along the time-axis to be processed in the backward direction. The initial values for the hidden state vectors are zero vectors $\overrightarrow{h}_{0} =  \overleftarrow{h}_{0}= \mathbf{{0}}$. After processing the whole sequence, we concatenate the final values of the GRU hidden state vectors. The concatenated final hidden state $h_{T}$ represents a summary of the input sequence. $h_{T}$ is projected through two fully connected layers to produce two vectors $\mu$ and $\hat{\sigma}$ which represent the mean and the logarithm of the variance of posterior distribution computed by the encoder. Each of $\mu$ and $\hat{\sigma} $ has a size of $N_z$. The log-variance output $\hat{\sigma}$  is converted into a non-negative standard deviation by the exponential operation.
\begin{equation}
\begin{split}
    h_{T} = [ \overrightarrow{h}_{T}; \;\; \overleftarrow{h}_{T}]\\
    \mu = \mathbf{W}_{\mu} h_{T} + b_{\mu} \\
    \hat{\sigma} = \mathbf{W}_{\sigma} h_{T} + b_{\sigma} \\
    \sigma = e^{\frac{\hat{\sigma}}{2}} \\
\end{split}
\end{equation}
Finally, we use the re-parameterization trick~\cite{kingma2013auto} to approximate the probabilistic sampling of the encoder output $\mathbf{z} \sim \mathcal{N}(\mu, \sigma^2) $ by a differentiable transformation defined upon $\mu$, $\sigma$ and an auxiliary random variable $\epsilon$.
 \begin{equation}
 \begin{split}
    \mathbf{z} = \mu + \sigma \odot \epsilon \quad \text{where } \epsilon \sim \mathcal{N}(0, \mathbf{I})
    \end{split}
\end{equation}

\subsubsection*{Decoder Design}
The decoder model $p_{\theta}(\mathbf{x} | \mathbf{z}, \mathbf{y}) $ translates the pair of latent space noise vector $\mathbf{z}$ and condition label $y$ into a time-series sequence of length $T$. As shown in Figure ~\ref{fig:decoder}, the decoder is an auto-regressive~\cite{graves2013generating} recurrent neural network that produces an output sequence one-step at a time. At each time-step, the generated output is fed back as an input into the next time step. Therefore, the decoder output at time step $t$ depends also on its own predictions at previous time-steps $<t $ in addition to the values of $\mathbf{z}$ and $\mathbf{y}$. 
Our decoder is built using a stack of three layers of gated recurrent units (GRU) neural networks. The vector $\mathbf{s}$ denotes the list of hidden states for the GRU units in the three layers and the vector $o^{(dec)}$ denotes the output of the last GRU layer. The initial state of the \textit{decoder} GRU units $\mathbf{s}_0$ is computed from the latent space vector $\mathbf{z}$ as in:
\begin{equation}
\begin{split}
\mathbf{s}_0 = tanh(\mathbf{W}_s \mathbf{z} + b_s) 
\end{split}
\end{equation}
At each time-step, the decoder takes its own generated value from the previous time-step $\bar{\mathbf{x}}_{t-1}$ along with the current hidden state value $\mathbf{s}_{t-1}$ to produce an output $o_{t}^{(dec)}$ and an updated hidden state $\mathbf{s}_{t}$. The last GRU layer output $o_{t}$ is projected through a fully connected layer to produce the final generated value $\bar{\mathbf{x}}_t \in \mathbb{R}^{N_d}$. We have found it useful to compute a context vector $\tilde{\mathbf{c}}$ by projecting both $\mathbf{z}$ and $\mathbf{y}$ through a fully connected layer and add this context vector the decoder input at each time-step. Using the context vector, $\tilde{\mathbf{c}}$, effectively adds a shortcut between the decoder output at each time-step and the encoder output $\mathbf{z}$ while processing long-sequences.

\begin{equation}
    \label{eqn:decoder}
    \begin{split}
        \tilde{\mathbf{c}} = \mathbf{W}_c [\mathbf{z} \; ; \mathbf{y}] + b_c \\
        o^{(dec)}_t, \mathbf{s}_t = Dec([\bar{\mathbf{x}}_{t-1} \;; \tilde{\mathbf{c}}], \mathbf{s}_{t-1}) \\
        \bar{\mathbf{x}}_{t} = \mathbf{W}_o \; o^{(dec)}_t + b_o
    \end{split}
\end{equation}
where the $Dec$ function represents the decoder 3-layered stack of GRUs.
\subsubsection*{Discriminator Design} 
The \textit{discriminator} $D$ is trained to distinguish between the samples produced by the \textit{decoder} when we feed random noise vectors into it and the examples drawn from the real dataset. Through the feedback it provides for the \textit{decoder} on the samples it generates, it forces the \textit{decoder} to improve the quality of its generated samples until the \textit{decoder} can produce samples that are sufficiently realistic to fool the \textit{discriminator} into accepting them as if they were drawn from the real dataset. As previously suggested in ~\cite{salimans2016improved}, we use a $L+1$ multi-class classifier instead of a binary classifier as our discriminator--where $L$ is the number of class labels and the $L+1$ class is the fake or ``generated`` class. Therefore, the goal of \textit{discriminator} is to classify all samples produced by the \textit{decoder} as label $L+1$ and classify samples drawn from the real data as their correct label $\mathbf{y} \in {1,...,L}$. The goal of the decoder while generating a new sample given a latent space vector $\mathbf{z}$ and a condition label $\mathbf{y} \in {1,..., L}$ is to fool the \textit{discriminator} into believing this sample is a genuine sample that belongs to the desired class label $\mathbf{y}$.

As such, this modification changes the original conditional GANs min-max equation, shown earlier in equation~\ref{eqn:cgan}, into the following:
\begin{equation}
    \label{eqn:discriminator}
    \begin{split}
\min_G \max_D V(D, G) =   \mathbb{E}_{\mathbf{x},\mathbf{y} \sim p_{\text{data}}}[\log D(\mathbf{x})_{\mathbf{y}} ]   \\
  + \mathbb{E}_{\substack{\mathbf{z} \sim p_{\mathbf{z}}(\mathbf{z})\\{ \mathbf{y} \sim Cat(\{1,..,L\})}}}[  \log \, D(G(\mathbf{z}, \mathbf{y}))_{L+1}  - \log \, D(G(\mathbf{z}, \mathbf{y}))_{\mathbf{y}}] \\
        \qquad \qquad \text{where, } D(\mathbf{x}) \in [0, 1]^{L+1} 
    \end{split}
\end{equation}

To realize the \textit{discriminator}, we use the 1-D convolutional classification model shown in Figure ~\ref{fig:encoder}. The \textit{discriminator} consists of three convolution layers followed by a final fully connected layer with a \texttt{softmax} output. Each convolution layer has 32 filters with filter size = $3$, and applied with stride = $3$ and zero padding. Additionally, we regard the output of the last convolution layer (layer 3) as an auxiliary output which is going to be useful to implement the \textit{feature matching} loss we describe in more details in Section ~\ref{sec:model_learning}.

\subsection{Model Learning}
\label{sec:model_learning}
The training of our model alternates between updating the \textit{discriminator} and updating both the VAE \textit{encoder} and \textit{decoder}. For simplicity, we will refer to the training of both the \textit{encoder} and the \textit{decoder} as the \textit{generator learning} because they are trained together--despite the fact that only the \textit{decoder} is needed for generating samples after the training has been finished. 

Further, the functions $\mathbf{E}(., \phi), \; \mathbf{G}(., \theta) \; \mathbf{D}(., \theta^d) $ are used to refer to the \textit{encoder}, \textit{decoder}, and \textit{discriminator}, respectively. Where the symbols $\phi$, $\theta$, and $\theta^d$ refer to the parameters of the three models, in the same order. 

\subsubsection*{Discriminator Learning}
The discriminator is trained to distinguish between the real data samples and those samples generated by the generator. It is a multi-class classifier with a $L+1$ probability distribution output where first $1 <= i <= L$ scores are the scores that the given sample is \textit{real} and predicted as class label $i$. The last score $L+1$ is reserved to represent the probability that the given sample is `\textit{fake}' or `\textit{generated}'. The adoption of a multi-class classifier instead of a binary classifier discriminator was proposed earlier by ~\cite{salimans2016improved} in the scope of semi-supervised learning to let the \textit{discriminator} provide a class-specific feedback signal to the generator outputs. Additionally, as noted by similar approach was used in ~\cite{chen2016infogan}, this forces the \textit{generator} to increase the mutual information between the GANs synthetic samples and the latent space condition codes. 
To train the discriminator, we sample batches of labeled examples from the training set, $\{(\mathbf{x}^{(i)}, \mathbf{y}^{(i)})\}_{i=1}^{M}$, and create a set of fake examples by feeding into the \textit{decoder} a set of randomly sampled pairs of latent space vectors and class label condition codes,  $\{(\mathbf{z}^{(i)}, \mathbf{y}^{(i)} )\}_{i=1}^{M}$. The objective of \textit{discriminator} learning is to minimize the following cost function:
\begin{equation}
    \label{eqn:disc_loss}
    \begin{split}
J_{disc}(\theta_d) =  &\; \mathbb{E}_{\mathbf{x},\mathbf{y} \sim p_{\text{data}}} \bigl[ - \log \mathbf{D}(\mathbf{x}; \theta^d)_{\mathbf{y}} \bigr]  \\
& +  \mathbb{E}_{\substack{\mathbf{z} \sim p_{\mathbf{z}}(\mathbf{z}) \\ \mathbf{y} \sim Cat(\{1,..,L\})}} \bigl[ -\log \mathbf{D}\bigl(\mathbf{G}(\mathbf{z}, \mathbf{y}; \theta); \theta^d\bigr)_{L+1} \bigr]
\end{split}
        \end{equation}
\subsubsection*{Generator Learning}
In the vanilla VAE model training, the training objective is defined based on the negative value of the evidence lower bound (ELBO), as shown in equation ~\ref{eqn:elbo}. Therefore, the original loss for VAE training is composed of two parts: the \textit{reconstruction error} objective 
and the enforcement of  smoothness on the latent space distribution of encoder outputs, making the encoder map examples into smooth regions in the latent space rather than single isolated points. This smoothness makes it more likely to produce realistic samples by feeding into the decoder values of latent space vectors sampled from the prior distribution $\mathbf{z} \sim p_{\mathbf{z}}(\mathbf{z})$, i.e.,  
\begin{equation}
\label{eqn:lvae}
    \mathcal{L}_{vae} = \mathcal{L}_{recon} + \mathcal{L}_{posterior}
\end{equation}
To improve the accuracy and diversity of the generated samples, we incorporate the discriminator feedback into the \textit{encoder}, and \textit{decoder} training objective by adding three more terms: feature matching $\mathcal{L}_{\text{features}}$, adversarial loss $\mathcal{L}_{adv}$, and the generator diversity $\mathcal{L}_{diverse}$. We explain the role and definition of each term as well as the equation of the total loss that combines them together.

\paragraph*{\textbf{Reconstruction loss $\mathcal{L}_{recon}$}} For any given single example from the training set $(\mathbf{x}^{(i)}, \mathbf{y}^{(i)})$, the reconstruction error measures how well the \textit{decoder} can recover the original input $\mathbf{x}^{(i)}$ after it has been compressed into the latent space code $\mathbf{z}$ produced by the \textit{encoder}. For sensor data readings with real values, we assume that the \textit{decoder} output is a Gaussian distribution with fixed variance. Therefore, the log-likelihood of \textit{decoder} output is proportional to the mean-squared error between the original reading value $\mathbf{x}^{(i)}$ and its own reconstruction through the autoencoder, i.e.,  $\bar{\mathbf{x}}^{(i)} = \mathbf{G}(\mathbf{E}(\mathbf{x}^{(i)}; \phi), \mathbf{y}^{(i)}; \theta)$. Therefore,
\begin{equation}
\label{eqn:lvae}
    \begin{split}
\mathcal{L}_{recon}(\mathbf{x}^{(i)}, \mathbf{y}^{(i)}; \phi, \theta) =   - \mathbb{E}_{\mathbf{z} \sim q_{\phi}(\mathbf{z} | \mathbf{x}^{(i)})} (\log \, p_{\theta}(\mathbf{x}^{(i)} | \mathbf{z}, \mathbf{y}^{(i)})) \\ 
\qquad \propto \left[ \frac{1}{T} \; \frac{1}{N_d} \,  \left\lVert \mathbf{x} - \mathbf{G} \bigl(\mathbf{E}(\mathbf{x}^{(i)}; \phi), \mathbf{y}^{(i)}; \theta\bigr) \right\rVert_2   \right] 
\end{split}
\end{equation}

\paragraph*{\textbf{Posterior loss $\mathcal{L}_{posterior}$}} the Kullback-Leibler Divergence loss $\mathcal{L}_{kl}$ in equation ~\ref{eqn:elbo} enforces a prior over the latent space distribution. When this prior distribution of latent space is selected to be IID Gaussian with zero mean and unit variance, the Kullback-Leibler divergence loss can be computed in the closed form~\cite{kingma2013auto}:
\begin{equation}
\label{eqn:recon}
\begin{split}
    D_{KL}\left(p(\mathbf{z}| \mu, \sigma^2) || \mathcal{N}(0, \mathbf{I}) \right) = & - \frac{1}{2} \frac{1}{N_z} (1 + \log \,\sigma^2 - \mu^2 - \sigma^2)
\end{split}
\end{equation}
where $\mu$ and $\hat{\sigma}$ are, respectively, the mean and log-variance of the posterior distribution outputted by the encoder network $q_{\phi}$, as we described in Section ~\ref{sec:model_structure}.

A common issue in VAE training is suffering from \textit{posterior collapse} where  the \textit{decoder}  ignores the latent space code $\mathbf{z}$.  As reported by previous research ~\cite{bowman2015generating,razavi2019preventing}, this is more likely to happen when the \textit{decoder} is by itself a powerful model such as the RNN \textit{decoder} in our case. The VAE \textit{posterior collapse} happens during the early steps of training when the model finds it is easier to bring down the KL-divergence component of equation ~\ref{eqn:lvae} rather than the reconstruction error. Therefore, $\mathcal{L}_{kl}$ goes rapidly down to nearly zero, after that the \textit{decoder} is optimized by itself to minimize the \textit{reconstruction error} while ignoring the \textit{encoder} output. Thus, there will be no gradient signal passed between the two models, i.e., the \textit{encoder} and \textit{decoder} have no influence on each other~\cite{bowman2015generating}. To address this issue, we use the `free bits'~\cite{kingma1606improving,razavi2019preventing} method that modifies VAE loss such that:
\begin{equation}
\label{eqn:posterior}
\mathcal{L}_{posterior}(\mathbf{x}, \mathbf{y}; \phi) = max\Bigl(D_{KL}\bigl(p(\mathbf{z}| \mu, \sigma^2) || \mathcal{N}(0, \mathbf{I}) \bigr) -\delta, 0\Bigr)
\end{equation}
\noindent where the KL-divergence is minimized only once until it surpasses a given threshold $\delta$--which we pick as $\delta=0.1$.
To ensure that the model learns how to pack useful information between the encoder and the decoder, we use a cost annealing scheme~\cite{bowman2015generating} that assigns a high weight to the \textit{reconstruction loss} and a nearly zero weight to the \textit{posterior loss} at the early steps of training. Then, we gradually and smoothly increase the weight of \textit{posterior loss} while decreasing the weight of the \textit{reconstruction loss}. This way encourages the model to pack useful information between the encoder and the decoder through the latent space code $\mathbf{z}$. The annealing scheme of the training cost is described in more details at the end of Section in equation ~\ref{eqn:annealing}.

\paragraph*{\textbf{Feature matching loss $\mathcal{L}_{feats}$}} The original reconstruction loss of the VAE training is based on the element-wise distance between the two vectors of the original input $\mathbf{x}^{(i)}$ and its reconstruction $\bar{\mathbf{x}}^{(i)}$ in the raw data space. This objective, however, leads to outputs that are blurry and lack fine details. Feature matching ~\cite{salimans2016improved} encourages the model to reduce the distance between the higher levels features of the original input $\psi(\mathbf{x}^{(i)})$ and those of its reconstruction $\psi(\bar{\mathbf{x}}^{(i)})$ . This encourages the model to maintain the holistic attributes of the data points while providing robustness against noise as well as in-variance against transformations such as signal shift.  The operator $\psi$ may be provided by either a domain-specific feature extraction algorithm or by taking the values of one of the hidden layers in a classifier. Since our \textit{discriminator} model is trained as a multi-class classifier to predict the correct label of input examples, we reuse the \textit{discriminator} as a feature extractor and, accordingly, the $\psi$ is chosen to be the output values of the last convolution layer in the \textit{multi-class discriminator} model. Therefore, the features reconstruction loss is defined as:
\begin{equation}
\label{eqn:features}
\begin{split}
    \mathcal{L}_{feats}(\mathbf{x}; \mathbf{y};\; \phi, \theta) = & \frac{1}{d_f} \left\lVert \psi(\mathbf{x}) - \psi(\bar{\mathbf{x})} \right\rVert_2 \\
     \quad \text{where, } & \bar{\mathbf{x}} = \mathbf{G}\bigl(\mathbf{E}(\mathbf{x};\; \phi\bigr), \mathbf{y};\; \theta)
\end{split}
\end{equation}
where $d_f$ is the dimension of the feature vector. I.e., $\psi(\mathbf{x}) \in \mathbb{R}^d$.
\paragraph*{\textbf{Adversarial Loss $\mathcal{L}_{adv}$}} The adversarial training loss of the \textit{decoder} is based on the feedback it receives from the \textit{discriminator} on its generated samples. This directs the \textit{generator (i.e. the Decoder)} to learn how to improve the quality of its generation by matching the class condition label code $\mathbf{y}$, i.e.,
\begin{equation}
\label{eqn:adv}
    \begin{split}
        \mathcal{L}_{adv}(\mathbf{z}, \mathbf{y}; \theta) = & - \log \mathbf{D}\bigl(\mathbf{G}(\mathbf{z}, \mathbf{y}; \; \theta); \; \theta_{d}\bigr)_{\mathbf{y}} \\
        \text{where } & \mathbf{z} \sim p_{\mathbf{z}}(\mathbf{z}) , \mathbf{y} \sim Cat(\{1,..,L\})
    \end{split}
\end{equation}
\paragraph*{\textbf{{Diversity loss $\mathcal{L}_{diverse}$}}} Training GANs requires finding the Nash equilibrium between two non-cooperating adversaries.  However, this process is known to unstable for training GANs as the \textit{discriminator} and \textit{generator} may train in orbits without convergence.  This is due to the fact that gradient-descent optimization is not well suited for the task of finding the Nash equilibrium. One common  symptom of GAN training failure is \textit{mode collapse}, where the \textit{generator} produces repeated samples that are essentially replicas of instances that were successful in fooling the discriminator. After the \textit{discriminator} identifies that these samples are \textit{fake}, the \textit{generator} will pick another mode to repeat, and so on. This prevents the \textit{generator} from producing samples with high diversity. This issue may be amplified in our model due to the \textit{posterior collapse} where the decoder may depend on itself as a powerful generative model and ignores the latent code $\mathbf{z}$ it receives from the encoder during the training. The \textit{diversity loss} penalizes this situation by the forcing the \textit{decoder} to utilize the latent code vector $\mathbf{z}$. To compute, the \textit{diversity loss}, we use the \textit{encoder} to reconstruct the latent code vector from samples generated by the decoder. The \textit{diversity loss} is defined as the mean-squared-error between the original latent code $\mathbf{z}$ and its reconstruction by the \textit{encoder} $\bar{\mathbf{z}}$. At the case when the \textit{decode} is suffering from `mode collapse`, it will ignore the latent code and produces identical samples. In such a case, the \textit{encoder} will be unable to recover the latent code from the samples, leading to a high penalty for the \textit{decoder}. The \textit{diversity loss} is defined as:
\begin{equation}
\label{eqn:diverse}
    \begin{split}
        \mathcal{L}_{diverse}(\mathbf{z}, \mathbf{y}; \; \phi, \theta) = & \left\lVert\mathbf{E}(\mathbf{G}(\mathbf{z}, \mathbf{y}; \theta); \phi) - \mathbf{z} \right\rVert_2 \\
      \text{where } & \mathbf{z} \sim p_{\mathbf{z}}(\mathbf{z}) , \mathbf{y} \sim Cat(\{1,..,L\})
    \end{split}
\end{equation}

\paragraph*{\textbf{Total training cost for generator}}
To summarize, the total training cost of the \textit{encoder} and \textit{decoder} models is defined as:
\begin{equation}
\label{eqn:gen_loss}
\begin{split}
J_{total}(\phi, \theta) =  \mathbb{E}_{\mathbf{x},\mathbf{y} \sim p_{\text{data}}} \Bigl[ & \eta_t \; \mathcal{L}_{recon}(\mathbf{x}, \mathbf{y}; \phi, \theta) \\
                                          & + \bigl(1-\eta_t\bigr) \bigl(\beta \mathcal{L}_{posterior}(\mathbf{x},\mathbf{y}; \phi) \\
                                         & +  \lambda_f \mathcal{L}_{feats}(\mathbf{x}, \mathbf{y}; \phi, \theta) \bigr)\Bigr]  \\
                                    \quad +   \mathbb{E}_{\substack{\mathbf{z} \sim p_{\mathbf{z}}(\mathbf{z}) \\ \mathbf{y} \sim Cat(\{1,..,L\})}} \Bigl[ & \bigl(1-\eta_t\bigr) \Bigl(\lambda_a \; \mathcal{L}_{adv}(\mathbf{z}, \mathbf{y}; \theta)  \\
 & + \lambda_{d}\; \mathcal{L}_{diverse}(\mathbf{z}, \mathbf{y}; \theta) \Bigr)\Bigr] 
\end{split}
\end{equation}
Where the $\beta=0.2, \lambda_f=1, \lambda_a=1, \lambda_d=0.2$ are weighting coefficients empirically chosen to balance the values of the different loss components. The $\eta_t$ is a decay function chosen to be the `inverse sigmoid decay`~\cite{bengio2015scheduled} with $k=200$.
\begin{equation}
\label{eqn:annealing}
    \begin{split}
        \eta_t = max \bigl( \frac{k}{k+\exp(t/k)}, 0.1 \bigr)
    \end{split}
\end{equation}
The goal of $\eta_t$ is to focus the training at the early step on only the \textit{reconstruction loss} $\mathcal{L}_{recon}$. Then gradually, add the other losses and decrease the importance of \textit{reconstruction loss}. We have empirically found this technique improves the stability of training. At the early steps, the output of the \textit{generator} will be too different from the real data and easy for the \textit{discriminator} to distinguish, leading to a saturation of the \textit{adversarial loss}.  We avoid this by focusing more on the \textit{reconstruction loss} at the early steps and then introduce the \textit{adversarial loss} after the model has started to produce sensible outputs. Also, the annealing scheme helps to avoid the \textit{posterior collapse} issue we discussed earlier in this Section. The gradual increase of the importance of the \textit{posterior loss}, $\mathcal{L}_{posterior}$, lets the model focuses first on using the latent space value $\mathbf{z}$ to store useful information in order to minimize the reconstruction loss then gradually starts to minimize the \textit{posterior loss}, $\mathcal{L}_{posterior}$, to match the latent space prior distribution.

The model is trained for 5,000 epochs using Adam optimizer with batch size = 256, and learning rate = 0.001.

\begin{algorithm}[!t] 
\caption{\name Model Training Algorithm} 
\label{alg:model_train} 
\begin{algorithmic}[1] 
    \REQUIRE a dataset of labeled training examples $\mathcal{D}^{train}=\{(\mathbf{x}^{(i)}, \mathbf{y}^{(i)}\}_{i=1}^{N}$.
    \STATE t = 0
    \STATE Initialize the weights of \textit{encoder} $\phi$, \textit{decoder} $\theta$, and \textit{discriminator} $\theta_d$ with random weights.
    \FOR{\texttt{number of training epochs}}
        \STATE $t = t+1$.
        \STATE Compute $\eta_t $ according to equation         ~\ref{eqn:annealing}.

        \FOR{\text{each batch} $B_d =\{(\mathbf{x}^{(i)}, \mathbf{y}^{(i)}) \}_{i=1}^{M}$ \text{in training data} $\mathcal{D}^{train}$}
            \STATE Sample a batch  of latent variables $B_{z,c} = \{(\mathbf{z} \sim p_{\mathbf{z}}(\mathbf{z}), \mathbf{y} \sim Cat(\{1,..,L\}))\}$.
            \STATE Use the training data batch $B_d$ and the latent variables batch $B_{z,c}$ to update the \textit{discriminator} weights $\theta_d$ to minimize the cost given in equation ~\ref{eqn:disc_loss}.
            \STATE Use the training data batch $B_d$ to compute equations ~\ref{eqn:recon}, ~\ref{eqn:posterior}, and ~\ref{eqn:features}.
            \STATE Sample another batch  of latent variables $B'_{z,c} = \{(\mathbf{z} \sim p_{\mathbf{z}}(\mathbf{z}), \mathbf{y} \sim Cat(\{1,..,L\}))\}$.
            \STATE Use $B'_{z,c}$ to compute equations ~\ref{eqn:adv}, and ~\ref{eqn:diverse}.
            \STATE Update the \textit{encoder} weights $\phi$ and \textit{decoder} weights $\theta$ to minimize the total cost given in equation ~\ref{eqn:gen_loss}.
            \ENDFOR
    \ENDFOR
\end{algorithmic}
\end{algorithm}

\subsection*{Model Summary}
\name consists of three different models: \textit{encoder}, \textit{decoder} and \textit{discriminator}. We extend the vanilla variational autoencoder training objective by including additional terms to improve the quality and diversity of generated samples. The procedure for training of \name is given in Algorithm ~\ref{alg:model_train}.

\section{Evaluation}
\label{sec:evaluation}
In the following section, we describe our experiments and evaluation results. We used two different datasets: ECG signal classification, and activity classification from motion sensors. The dataset are described in details in section~\ref{sec:eval_datasets}. We compare the quality of the synthetic data generated by our model against different baselines, which are described in section~\ref{sec:eval_baseline}. Choosing the right metric for evaluating the quality of generative models is still an active topic of research. In addition to the visual quality of samples, most of metrics currently in use are specific to the kind of data being generated such as the Inception Score~\cite{salimans2016improved} and The Fr\'echet Inception distance (FID)~\cite{heusel2017gans} of image generation, and the perplexity score~\cite{chen1998evaluation} for text generation. In our paper, we evaluate the quality of generated sensor data using metrics that measure their conditional generation accuracy score  (Section ~\ref{sec:eval_accuracy}), diversity  of samples score (Section ~\ref{sec:eval_diversity}), and novelty of samples (Section ~\ref{sec:eval_novelty}). Besides, we use two application-specific metrics of the overall quality of the generated data. The first application-specific evaluation measures the \textit{synthetic dataset utility} (Section ~\ref{sec:eval_utility}) which measures how well suited are the \textit{generated samples} to be used for training classification models that are evaluated on \textit{real} samples test data. The second application-specific evaluation is based on using the generative model to perform data imputation (Section ~\ref{sec:eval_imputation}) by filling in missing segments of sensor readings.

\begin{figure*}[!t]
\centering
\includegraphics[width=0.8\textwidth]{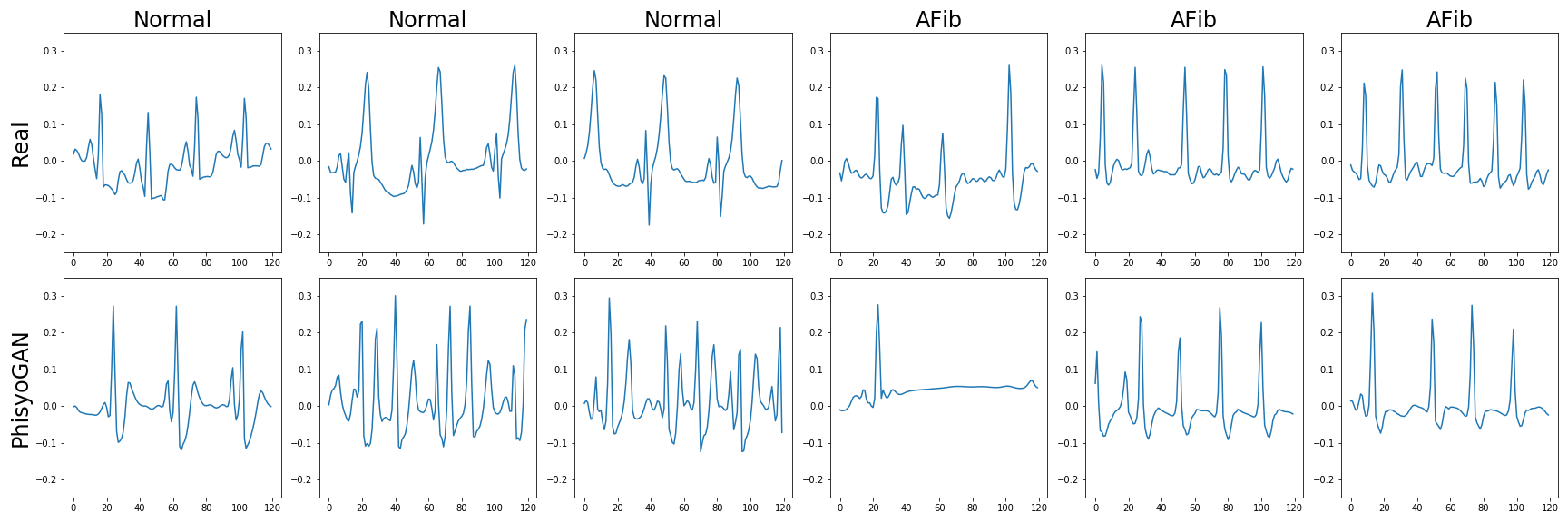}
\caption{Random samples of the real data (top row) and synthetic data (bottom row) generated by \name on the AFib classification ECG dataset. The title of each column indicate the class label of the samples. More samples and illustration of samples produces by other baseline models can be found in the supplemental material.}
\label{fig:result_ecg}
\end{figure*}

\begin{figure*}[!t]
\centering
\includegraphics[width=0.8\textwidth]{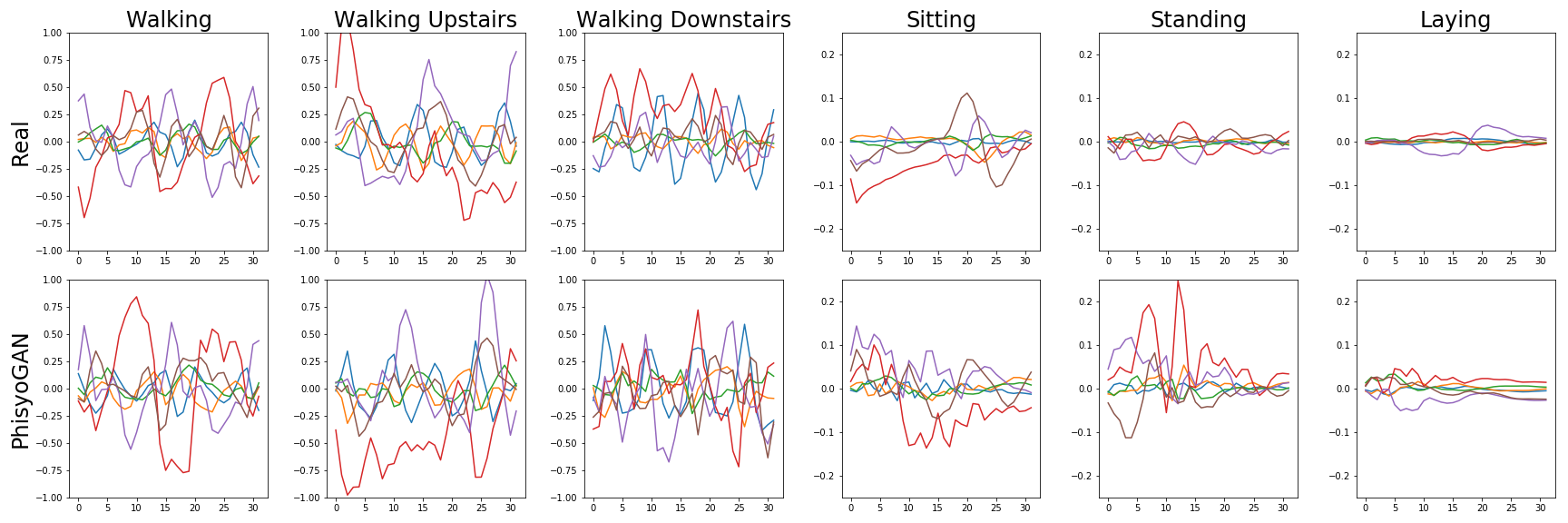}
\caption{Random samples of the real data (top row) and synthetic data (bottom row) generated by \name on the HAR dataset. The title of each column indicate the class label of the samples. More samples and illustration of samples produces by other baseline models can be found in the supplemental material.}
\label{fig:result_har}
\end{figure*}

\subsection{Datasets}
\label{sec:eval_datasets}

\subsubsection*{ECG dataset for AFib classification:} Atrial Fibrillation (AFib) is an irregular heartbeat (arrhythmia) disorder. It is considered the most common arrhythmia type, occurring in 1-2\% of the general world population and leads to a significant increase in the risks of death, strokes, hospitalization, and heart failure~\cite{developed2010guidelines}. The Electrocardiography (ECG) signal is considered the most common method for arrhythmia classification. The arrhythmia detection dataset of ~\cite{yildirim2018arrhythmia} contains 1000 fragments of ECG signals from 45 persons taken from the MIT-BIH Arrhythmia Database. The dataset fragments correspond to 17 classes including the \textit{Normal Sinus Rhytm} (NSR), the \textit{Atrial Fibrillation} (AFib) and 15 others. In our study, we focus on learning how to generate only (NSR) and AFib classes because they are the most important rhythms and because the other rhythms had a significantly smaller number of examples in the dataset. Therefore, we use the  418 examples that correspond to NSR and AFib classes. Each sample corresponds to 10 seconds of ECG samples recorded at 300 Hz sampling rate. For computational efficiency, we sub-sample the signal to 30 Hz. We split the data into train and test subsets using 75\%, 25\% split ratios. We train a recurrent neural network-based classification model on the dataset, after sub-sampling, which achieves 97.14\% classification accuracy, which is on-par with the state-of-the-art classification accuracy reported by ~\cite{yildirim2018arrhythmia}. Random samples of the dataset are shown in the top row of Figure ~\ref{fig:result_ecg}.
\subsubsection*{Motion Sensors for Human Activity Recognition} Human activity recognition is important for many reasons, such as elderly fall detection~\cite{chen2006wearable} and the assessment of Parkinson disease patients~\cite{aghanavesi2019motion}. Activity recognition in wearable devices relies on using the embedded motion sensors such as accelerometer and gyroscope. The UCI \textit{human activity recognition} HAR dataset~\cite{anguita2013public} includes 10,299 examples collected from smartphone attached to the waist of 30 volunteers while performing six different activities: walking, walking upstairs, walking downstairs, sitting, standing, and laying. The dataset is split into training and test sets using 70\%, 30\% split ratios. We trained a recurrent neural network-based classification model on the dataset, which achieves 89.74\% test set classification accuracy. We use this dataset as an example for learning how to generate a \textit{multi-class} and \textit{multi-dimensional} (each time step has six values corresponding to the $X$, $Y$, $Z$ axis values for each of the accelerometer and gyroscope sensors) time-series data. Samples of the dataset are shown in the top row of Figure ~\ref{fig:result_har}.

\noindent Notably, the dataset we have used in our experiments is considered more challenging than those used in the original experiments of previous work. For example, ~\cite{alzantot2017sensegen} did not perform conditional generation or utility evaluation. ~\cite{esteban2017real} performed experiments on only toy dataset (e.g., sin-waves) and short low-frequency frequency simple classification tasks. Likewise, ~\cite{wang2018sensorygans} only studies the generation of accelerometer data while considering only highly dissimilar classes (e.g., only \textit{walking} vs. \textit{standing}) while we consider the more challenging case of generating six classes including classes are highly similar to each other (e.g., \textit{walking}, \textit{walking upstairs}, and \textit{walking downstairs}).
\subsection{Baseline algorithms}
\label{sec:eval_baseline}
In comparison to the vast amount of research done on image and text generation, significantly less success has been made towards the conditional generation of high-quality synthetic sensor dataset. Among the notable efforts in this space that we are aware of is the work of ~\cite{alzantot2017sensegen} trains a maximum-likelihood based recurrent neural network for \textit{unconditional} generation of accelerometer sensor readings. Both the work of ~\cite{esteban2017real} and ~\cite{wang2018sensorygans} uses adversarial training to train a recurrent neural network of producing real-valued time-series values. We include those methods as baselines and compare their performances against PhisyoGAN according to the measures of condition accuracy, diversity, and novelty and synthetic data utility.

We use the following baseline models:

\begin{itemize}
    \item \textbf{CRNN} \textit{(Conditional RNN-Model)}: This is an auto-regressive~\cite{graves2013generating} recurrent neural network model similar which is trained by maximizing the likelihood of predicting each next time-step values given the previous values. This baseline can be considered as an extension to ~\cite{alzantot2017sensegen} with the additional support of performing conditional generation which was achieved by conditioning the RNN at each step $t$ on the latent noise vector, the class condition label and the prediction output at previous time step, $[\mathbf{z}, \mathbf{y}, \tilde{\mathbf{x}}_{t-1}]$,  in the same way as our decoder introduced earlier in equation ~\ref{eqn:decoder}.
    
    \item \textbf{CVRAE} \textit{(Conditional Variational Recurrent Auto-Encoder)}: This is a conditional variational autoencoder with a recurrent encoder and recurrent decoder. The architecture of \textit{encoder} and \textit{decoder} were same as those used in our model. But the training objective is different. The training objective used for \textit{CVRAE} is the vanilla conditional VAE training loss introduced earlier in ~\ref{eqn:lvae}.
    \item \textbf{RCGAN} \textit{(Recurrent Conditional GAN)}: This model mimics the structure and training method of the conditional recurrent generative model introduced in ~\cite{esteban2017real}. It consists of a recurrent neural network generator which is trained with the GANs training objective shown in equation ~\ref{eqn:discriminator}. Notably, this model is not auto-regressive and the RNN input at each time-step dependent consists of only the latent space code and class condition label, i.e. $[\mathbf{z}, \mathbf{y}]$, but not on the previous predictions made by the generator.
    \item \textbf{RCGAN-AR} \textit{(Recurrent Conditional GAN-Auto-regressive)}: This baseline extends the \textbf{RCGAN} model by introducing feedback connections that go from the generator output at one time-step to its input in the next time-step. Therefore, the generator behaves exactly like our decoder, shown in equation ~\ref{eqn:decoder}, but it is still trained with same training objective as the RCGAN model. We include this baseline because we notice that RCGAN, due to its lack of auto-regressive feedback connection, fails to produce samples with a length exceeding the length of training examples.
\end{itemize}

To ensure a fair comparison between all models, the size of \textit{CRNN}, \textit{RCGAN}, \textit{RCGAN-AR} is identical for the size of our \textit{decoder}, described earlier in Section~\ref{sec:model_structure}. While the baseline \textit{CVRAE} has the same architecture of its encoder and decoder as the encoder and decoder of \name. Also, models that required a \textit{discriminator} for adversarial training (\textit{RCGAN}, \textit{RCGAN-AR}) were trained using the same multi-class convolutional discriminator that used to train \name, which we described earlier in Section~\ref{sec:model_structure}. Therefore, the five models are only different in their model training technique.

\subsection{Evaluation Results}
\label{sec:eval_results}

Figures ~\ref{fig:result_ecg}, and ~\ref{fig:result_har} provides a visual illustration of randomly selected samples from the real data and randomly selected samples of the synthetic data generated by \name for each class of the ECG AFib classification and the human activity recognition (HAR) classification datasets, respectively. 
More samples and illustration of samples produces by other baseline models can be found in the supplemental material. Since it is more challenging to rely on visual inspection of sensor measurements than images and text as an evaluation metric. Therefore, in the rest of this section, we introduce test results that measure the different aspects of synthetic data quality: accuracy, diversity, and novelty of the synthetic samples. In addition to the aforementioned evaluation criteria, We also conduct additional task-specific metrics: the \textit{synthetic dataset utility}, and the \textit{data imputation quality}.

\subsection{Conditional Generation Accuracy Score}
\label{sec:eval_accuracy}

\begin{table}[!h]
    \centering
    \renewcommand{\arraystretch}{1.5}
    \begin{tabular}{lccc}
    \toprule
    Model &  HAR Dataset &  & AFib Dataset \\
    \midrule
    Real Data &  89\% &  & 97\%  \\
    \cdashline{1-4}
    CRNN  & 76.0\%  & & 67\%  \\
    CVRAE & 72.0\% &  & 67\%  \\
    RCGAN &  100\% &  & 100\% \\
    RCGAN-AR & 82\% &  & 100\% \\
    \name & 90\% &  & 94\%  \\
    \bottomrule
    \end{tabular}
    \caption{The \textit{conditional generation score} of synthetic dataset produced by each generative model. The first row indicates the accuracy of the \textit{oracle} model which is trained a training dataset from the real data.}
    \label{tab:conditional_score}
\end{table}

The \textit{conditional generation accuracy score} evaluates the rate by which the generated sensor readings match the class label that the generator was conditioned upon to generate those samples.  To compute the score value, for each dataset, we train a high accuracy model on the \textit{real dataset} and use this model as a \textit{oracle} that predicts a classification label of each synthetic sample. To evaluate each generative model, we generate a large set of synthetic samples (with size = 10 times the size of the real training data) produced by that model and use the \textit{Oracle} model to predict a label for those samples. The rate by which the \textit{Orcale} predictions matches the \textit{condition} code of the generated samples represents the accuracy of conditional generation. The results of \textit{conditional generation score} are shown in ~\ref{tab:conditional_score}.

The result from Table ~\ref{tab:conditional_score} that the models trained with adversarial training (i.e., RCGAN, RCGAN-AR, and \name) have a significantly higher \textit{conditional generation score} than models trained with maximum-likelihood (i.e., CRNN, and CVRAE). This indicates that adversarial-trained generative models are more likely to produce samples that will look, according to the \textit{oracle} model, as the class they were supposed to match. However, the \textit{conditional generation score} is not a sufficient metric to assess the quality of the generative models because it does not assess the \textit{intra-class} diversity of generated samples. Neither, it does evaluate the novelty of the generated samples to inspect whether or not the generative model is memorizing samples from training data. Therefore, we introduce two other metrics: the \textit{diversity} and \textit{novelty} scores.

\subsection{Diversity of Samples Score}
\label{sec:eval_diversity}
\begin{table}[!ht]
\renewcommand\arraystretch{1.5}
    \centering
    \begin{tabular}{lccc}
    \toprule
    Model & HAR Dataset & & AFib Dataset \\
    \midrule
    Real data  & 1.00 & & 1.00 \\
    \cdashline{1-4}
    CRNN  & 0.43 &  & 1.03  \\
    CVRAE & 0.38 &  & 1.01 \\
    RCGAN &  0.27 &   & 0.06   \\
    RCGAN-AR & 0.23 &  & 0.07   \\
    \name &  1.14 &  & 0.87  \\
    \bottomrule
    \end{tabular}
    \caption{Diversity Scores of Synthetic Datasets Generated by Different Generative Models}
    \label{tab:diversity}
\end{table}
Mode collapse is a common pitfall for GANs~\cite{theis2015note}. It is defined by the case when the \textit{generator} produce synthetic samples that are very similar to each other. On the other hand, we want the \textit{generator} to produce samples that are accurate, diverse, and novel from those in the training dataset.  In previous research, ~\cite{wang2018sentigan} defined a score metric to evaluate the diversity of generated text. In the same way, we define a \textit{diversity score} of the synthetic dataset according to the equation in ~\ref{eqn:diversity}.
\begin{equation}
    \label{eqn:diversity}
    \begin{split}
        Diversity \bigl( \mathcal{S}^{(i)} \bigr)  = & \frac{1}{\Lambda} \; min \; \bigl\{ \texttt{DTW} (\mathcal{S}^{(i)}, \mathcal{S}^{(j)})  \bigr\}_{j=1}^{j=|\mathcal{S}|, j\neq i} \\
        \text{Where, } \Lambda=  & \frac{1}{|\mathcal{D}|} \sum_{i=1}^{i=|\mathcal{D}|} { min \; \bigl\{ \texttt{DTW} (\mathcal{D}^{(i)}, \mathcal{D}^{(j)})  \bigr\}_{j=1}^{j=|\mathcal{D}|, j\neq i}}
    \end{split}
\end{equation}
Given a set of synthetic examples $\mathcal{S}$ and another set of real data examples $\mathcal{D}$, the \textit{diversity score} of an individual sample $S^{(i)}$ is defined as the distance between the $i$\textsuperscript{th} sample and its nearest neighbor in the synthetic dataset $\mathcal{S}$. Distances are measured using the dynamic time warping (DTW)~\cite{salvador2007toward} distance measure, which is a reliable measure of time-series dissimilarity due to robustness to minor translations and variations. The \textit{diversity score} of the whole dataset $\mathcal{S}$ is the average of the \textit{diversity score} assigned for individual samples. In order to have a normalized score value where, as a reference, the \textit{diversity score} of the original real dataset $\mathcal{D}$ is equal to 1 we divide the \textit{diversity score} of each sample in a synthetic dataset by the normalizer $\Lambda$ which is the average dynamic-time-warping distance between each example from the real dataset $\mathcal{D}$ and its nearest neighbor from the same dataset.

 The results for computing the diversity scores on the synthetic datasets produced by \name and the baseline generative models are shown in Table ~\ref{tab:diversity}. The result shows that \name has a significantly higher diversity of generated samples than the other methods trained with adversarial training RCGAN, and RCGAN-AR which reflects how \name had much less mode collapse then the other models that relied only on the vanilla \textit{adversarial training} objective.

\subsection{Novelty of Samples Score}
\label{sec:eval_novelty}

\begin{table}[!h]
\renewcommand\arraystretch{1.5}
  \centering\small
    \begin{tabular}{lcc}
    \toprule
    Model & HAR Dataset & AFib Dataset \\
    \midrule
    CRNN  & 1.09 & 1.33   \\
        CVRAE & 1.33 & 1.52   \\
    RCGAN &  1.67 & 1.15   \\
    RCGAN-AR & 1.75 & 1.00   \\
    \name & 1.35 & 1.02  \\
    \bottomrule
    \end{tabular}
    \caption{Novelity Scores of Synthetic Datasets Generated by Different Generative Models}
    \label{tab:novelty}
\end{table}
Generative models are desired to learn the underlying distribution of the training dataset and produce samples that are both \textit{novel} and \textit{realistic} rather than over-fitting the real dataset set. We extend the idea of ~\cite{wang2018sentigan} to evaluate the novelty of the synthetic samples generated by the models under our study using the \textit{novelty score} shown in ~\ref{eqn:novelty}. 
\begin{equation}
    \label{eqn:novelty}
    \begin{split}
        Novelty \bigl( \mathcal{S}^{(i)} \bigr)  = & \frac{1}{\Lambda} \; min \; \bigl\{ \texttt{DTW} (\mathcal{S}^{(i)}, \mathcal{D}^{(j)})  \bigr\}_{j=1}^{j=|\mathcal{D}|} \\
        \text{Where, } \Lambda =  & \frac{1}{|\mathcal{D}|} \sum_{i=1}^{i=|\mathcal{D}|} { min \; \bigl\{ \texttt{DTW} (\mathcal{D}^{(i)}, \mathcal{D}^{(j)})  \bigr\}_{j=1}^{j=|\mathcal{D}|, j\neq i}}
    \end{split}
\end{equation}
Given a dataset of synthetic samples $\mathcal{S}$ and the training dataset $\mathcal{D}$ that was used to train the model which produced $\mathcal{S}$, the novelty score of an individual sample $\mathcal{S}^{(i)}$ is measured as its distance to the nearest neighbor from samples in $\mathcal{D}$. Distances are measured using the dynamic time warping ~\cite{salvador2007toward}. Novelty scores are also normalized by dividing their value upon the average distance to nearest neighbor between samples in $\mathcal{D}$ and each other. Therefore, a novelty score equal to zero indicates that the \textit{synthetic} sample is a replica of another sample in the training set. When the novelty score of a \textit{synthetic sample} is equal to one, this indicates that the sample is, on average, as close to samples in the training data as samples from the training data are close to each other. The novelty score of the entire synthetic dataset $\mathcal{S}$ is the mean value of novelty score for each individual sample in $\mathcal{S}$.

\begin{figure}[!htp]
    \centering
    \includegraphics[width=0.4\textwidth]{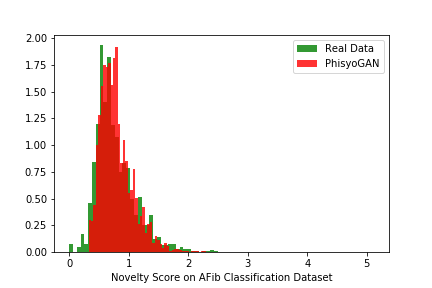}
    \caption{Distribution of Novelty Score on AFib dataset}
    \label{fig:novelty_hist}
\end{figure}

The novelty score of each \name and the other baseline generative model training methods are shown in Table ~\ref{tab:novelty}. We notice that \name has a slightly lower \textit{novelty score} than other methods which may indicate that samples produced by \name are more similar to the training data samples. To investigate whether or not \name is memorizing samples from the training data, we compare the distribution of novelty scores for individual synthetic samples produced by \name on each dataset and the distribution of \textit{novelty scores} for real data in Figure ~\ref{fig:novelty_hist}. From the figure, we conclude that \name is not memorizing the training data samples because the probability density of novelty scores of synthetic samples by \name has very low value around zero.
\begin{table*}[!htp]
    \centering
\renewcommand\arraystretch{1.5}
    \begin{tabular}{lccccc}
    \toprule
    \multicolumn{1}{}{}& \multicolumn{2}{c}{HAR  Dataset} &  &
    \multicolumn{2}{c}{AFib Dataset} \\
    \cmidrule{2-3}\cmidrule{5-6}
    Training Dataset & RNN & SVM & &  RNN & SVM   \\
    \midrule
    \multicolumn{1}{l}{Real Data} & 89.0\% & 83.4\% & &  0.96 (97.0\%) & 0.77 (85.0\%)  \\
    \cdashline{1-6}
      \multicolumn{1}{l}{CRNN} & 32.4\%  & 35.9\% & & 0.46 (54.2\%) & 0.441 (38.0\%)   \\
      \multicolumn{1}{l}{CVRAE} & 46.3\% & 43.7\%&  &  0.50 (48.7\%)  & 0.44 (42.5\%)  \\
      \multicolumn{1}{l}{RCGAN} & 30.5\% & 37.6\% & & 0.52 (55.2\%) & 0.5 (66.6\%)\\
    \multicolumn{1}{l}{RCGAN-AR} & 29.1\%& 26.9\%  & & 0.48 (50.4\%) &  0.5 (66.6\%)  \\
      \multicolumn{1}{l}{\name} & \textbf{77.8\%} & \textbf{65.1\%} &  & \textbf{0.87 (88.6\%)} & \textbf{0.67 (77.8\%)} \\
    \bottomrule
    \end{tabular}
    \caption{Accuracy scores for classification models trained on \textit{synthetic} datasets generated by different generative models. The first row indicates the accuracy of the same model when trained on real data. For the AFib dataset, we use the area-under-the curve (AUC) score because the test dataset is highly imbalanced and show the accuracy between parentheses.}
    \label{tab:utility}
\end{table*}

\subsection{Utility of Synthetic datasets}
\label{sec:eval_utility}

Since datasets of physiological and medical sensor readings are often considered privacy-sensitive, laws and regulations impose a lot of constraints on how this data can be shared. This introduces a challenge for research teams who collect datasets and are willing to share it with other researchers or the public audience. As an alternative, those researchers may resort into generating \textit{synthetic dataset} that does not belong to real patients but are produced with a \textit{generative model} trained on real patients data. The utility of synthetic dataset under this situation is reflected by how good are they to be used in lieu of the real data in the downstream task (commonly a classification task). 

Rather than evaluating the quality of a conditional generative model based on measuring the aspects of \textit{generation accuracy}, \textit{diversity}, and \textit{novelty}. An alternative way is to measure how good are they to produce data suitable for a downstream task. Even though the downstream is generally unknown at the model training time, a good generative model that learns the underlying distribution of the data should be able to produce synthetic data that are as good to be used in any downstream task as the training real data.  Following this approach, a recently proposed metric for the quality of conditional generative models is the \textit{classification accuracy score}~\cite{ravuri2019classification}. The \textit{classification accuracy score} evaluates conditional generative models by training a classification model using \textit{only} synthetic data and validates the accuracy of the trained model on real test data. This idea was also proposed earlier in ~\cite{esteban2017real} under the name of `TSTR` (Train on Synthetic Test on Real). In their recent study on ImageNet generative models, ~\cite{ravuri2019classification} have noticed that classification models trained on a dataset of synthetic images generated by the state-of-the-art BigGAN~\cite{brock2018large}, will have a top-1 and top-5 classification accuracy that are 27.9\% and 41.6\% less than the accuracy scores of models trained on real data. It was also observed that \textit{classification accuracy score} for a variety of generative models does not correlate with other metrics such as the Inception score~\cite{salimans2016improved}, and FID~\cite{heusel2017gans} which indicates the challenging nature of how to evaluate generative models.

\begin{table*}[!t]
\renewcommand{\arraystretch}{1}
    \centering
    \begin{tabular}{lccccccccccc}
   \toprule
   \multicolumn{1}{l}{}&\multicolumn{5}{c}{AFIB} & & \multicolumn{5}{c}{HAR}\\
   \cmidrule{2-6} \cmidrule{8-12}
    \multicolumn{1}{l}{} & \multicolumn{2}{c} {{MCAR}} & & \multicolumn{2}{c}{Missing Segment} & & \multicolumn{2}{c} {{MCAR}} & & \multicolumn{2}{c}{Missing Segment} \\
    \cmidrule{2-3} \cmidrule{5-6} \cmidrule{8-9} \cmidrule{11-12}
     \multicolumn{1}{l}{} & MAE & Semantic Repair & & MAE & Semantic Repair & & MAE & Semantic Repair & & MAE & Semantic Repair \\
    \midrule
     \multicolumn{1}{l}{KNN} &  0.045 & 68.7\% &  & \textbf{0.051}  & 81.2\%          &  &0.063  & 98.6\% & & 0.077& 96.4\%   \\
    \multicolumn{1}{l}{MICE} &0.027 &  90.6\% &   & 0.056& 84.4\% &   & \textbf{0.036} &  95.1\%& & \textbf{0.076} & 82.6\%\\
\multicolumn{1}{l}{BRITS} &  \textbf{0.018}& \textbf{96.9\%} &  &  0.052    &   81.2\%         &  & 0.098 & 93.1\% & & 0.102& 85.3\%\\
     \multicolumn{1}{l}{\name} & 0.045 & 81.2\% &  & 0.067 & \textbf{90.6\%}         &  & 0.087 & \textbf{100\%} & &0.114 & \textbf{100\%} \\
    \bottomrule
    \end{tabular}
    \caption{Scores of Sensor Data Imputation}
    \label{tab:imputation_result}
\end{table*}

We train four different classification models, two for each task, on synthetic datasets. The first model is a deep-recurrent neural network-based model, while the second model is a traditional SVM classification model with human-engineered features, selected from best performing work in the literature on each task. To train models on synthetic data, we sample a set of \textit{generator} results with size = 10 times the size of the real training dataset. We have noticed that increasing the size of synthetic data will increase the test accuracy. Results for training those models on both the real data and synthetic datasets are shown in Table \ref{tab:utility}. The top row indicates the accuracy of the models trained on real data for each task. The rows of generative models show the accuracy of classification models trained by their synthetic datasets when evaluated on the \textit{real test dataset}.  The results in Table ~\ref{tab:utility} show that \name have significantly higher utility than the other generative model training methods. On HAR classification and the AFib classification tasks, classification models had only roughly 20\% and 10\% decrease in their accuracies when trained on synthetic dataset produced by \name instead of the real data. Models trained on synthetic datasets by other models had a much larger drop in their classification accuracies as shown in ~\ref{tab:utility}.  The high utility of synthetic datasets by \name reflects  their good balance between their \textit{generation accuracy} and \textit{diversity}. %

\subsection{Sensor Data Imputation}
\label{sec:eval_imputation}

Missing values in time-series sensor readings are pretty common due to environmental noise, e.g., wireless connection drops or a displaced sensor due to human body movements. These discontinuities significantly degrated the quality of the collected signals. Given that \name learns how to reconstruct and generate sensor data samples, we also investigate how \name can be used to \textit{repair} corrupted samples by imputing their missing values. We simulate two different scenarios of missing values:
\begin{itemize}[noitemsep,topsep=0pt,parsep=0pt,partopsep=0pt]
\item \textbf{Missing-Completely-at-Random (MCAR)}: In the \textit{missing at random} scenario, each time step of sensor readings is subject to be missing with 25\% dropout chance independent from the readings of other time steps.
\item \textbf{Missing Segment:} In the \textit{missing segment} scenario, a continuous segment with length = 25\% of the whole input example is removed. The starting position of the missing segment is selected uniformly between the start and 75\% of the original example length.
\end{itemize}

Given that an input example $\mathbf{x} \in \mathbb{R}^{T \times N_d}$ has been corrupted by missing some values, the mask variable $\mathbf{m} \in [0,1]^{T}$ is an array of indicator variables that represents whether a sensor readings value is present or missing. The corrupted signal $\mathbf{x}m$ is simply the hadamard, element-wise product between $\mathbf{x}$ and $\mathbf{x}m$ with broadcasting across the last dimension, i.e.,

\begin{equation}
    \mathbf{x}m = \mathbf{x} \odot \mathbf{m}
\end{equation}

 The goal of data imputation is to recover $\mathbf{x}$ from  $\mathbf{x}m$, $\mathbf{m}$, and $\mathbf{y}$. We utilize \name to impute the missing values by taking advantage of both the \textit{encoder}'s and \textit{decoder}'s abilities to reconstruct signals in a semantically preserving way. Without any change in \name's model training, we utilize it as an imputer in the following way.
 
 We first encode the incomplete sensor reading using the \textit{encoder} model, $\mathbf{E}$, into a latent space vector $\mathbf{z}$.
 \begin{equation}
     \begin{split}
         \mathbf{z} = \mathbf{E}(\mathbf{x}m; \phi) \\
    \end{split}
 \end{equation}
 Then, we utilize the \textit{decoder} to reconstruct the signal conditioned on the combined latent codes of $[\mathbf{z}; \mathbf{y}]$ like we illustrated earlier in Equation ~\ref{eqn:decoder} under the slight modification (which we applied only while using the model for imputation but not during the training):
\begin{equation}
    \label{eqn:decoder_impute}
    \bar{\mathbf{x}}_{t} =
    \begin{cases}
    \mathbf{W}_o \; o^{(dec)}_t + b_o, 
      & \text{if}\ \mathbf{m}_t=0 \\
      \mathbf{x}_t,  & \text{if}\ \mathbf{m}_t=1
    \end{cases}
\end{equation}
\begin{figure*}[!t]
  \centering
  \includegraphics[width=\textwidth]{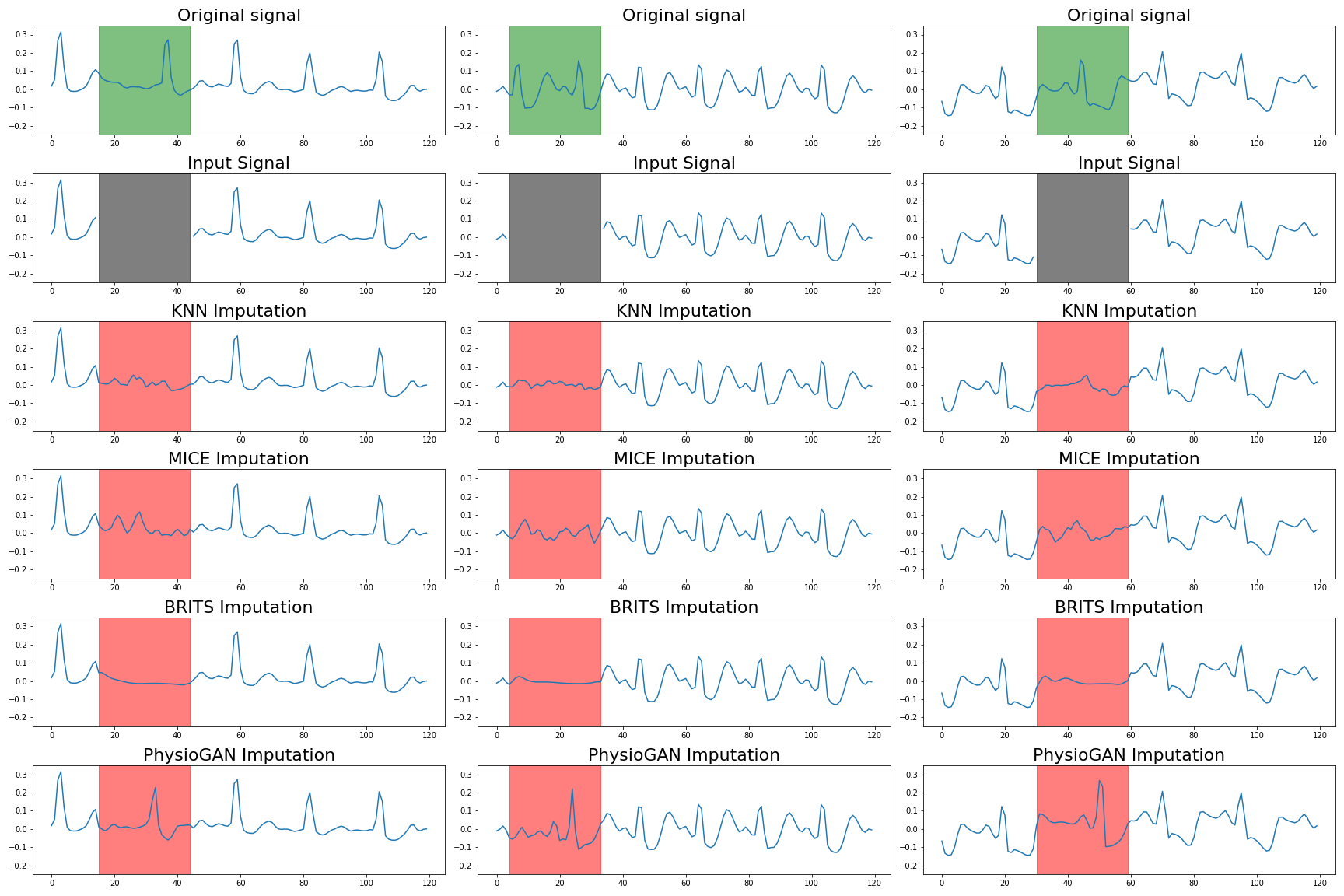}
    \caption{Different examples of using our model to fill in the 25\% missing segment in ECG data. The top row shows the original \textit{complete} data, the second row shows the input data with missing segments highlighted with a black background, the bottom row shows the result of using our model to fill in the missing segments.}
  \label{fig:imputation_result}
  \end{figure*}

To evaluate the performance of \name for data imputation, We simulate the two scenarios of missing values - missing at random, and missing segment - on 105 and 500 examples from the AFib and HAR datasets, respectively. In each scenario 25\% of sensor readings were considered missing. We compare \physgan performance on data imputation against the following baselines:

\begin{itemize}[noitemsep,topsep=0pt,parsep=0pt,partopsep=0pt]
    \item \textbf{KNN~\cite{friedman2001elements}:} The KNN selects the most similar neighbours from the training examples, \textit{with the same label $y$}, according to the \textit{euclidean distance} between their observed values. Then, it replaces the missing values with the mean of the corresponding values in the selected neighbours.
    \item \textbf{MICE:~\cite{azur2011multiple}} Multiple Imputation by Chained Equations (MICE)~\cite{azur2011multiple} is a widely used imputation method. It multiple imputations with chained
equations to fill missing values.
    \item \textbf{BRITS:~\cite{cao2018brits}}: BRITS~\cite{cao2018brits} is the state-of-the-art method for time-series imputation using neural networks. It uses a bidirectional recurrent neural networks. It treats missing values as variables in the bidirectional RNN computation graph. It uses delayed gradients in both forward and backward directions to learn how to predict the missing values. Together with learning to predict the missing values, the bidirectional RNN is also trained as classification model to improve she semantics of imputation.
\end{itemize}

Figure ~\ref{fig:imputation_result} shows a set of randomly selected examples of imputation results by different methods on the AFib dataset under the missing segment scenario. Additionally, we use the following metrics to compare the imputation results:

\begin{itemize}[noitemsep,topsep=0pt,parsep=0pt,partopsep=0pt]
    \item \textbf{Mean absolute error (MAE)}: The mean absolute error between the original and imputed signals as computed as
    \begin{equation*}
        MAE(\mathbf{x}, \bar{\mathbf{x}}) = \frac{1}{T \times d} \sum_{t=1}^{T}{\sum_{j=1}^{d}}{\vert \mathbf{x}_{t,j} - \bar{\mathbf{x}}_{t,j} \vert}
    \end{equation*}
    
    \item \textbf{Semantic Repair (SR)}: Since distribution of sensor readings is multi-modal, there might be more than one realistic imputation results. However, the MAE error metric will produce low error value for only one of them.  Additionally, the MAE metric is not sufficient to capture the change in the sensor data semantics after it has been imputed. Therefore, we propose the \textit{semantic repair} which, intuitively, measures how well the imputation method reduces the gap between classification accuracies of the \textit{complete} and \textit{incomplete} datasets.
    For a given dataset of original (complete) sensor readings $\mathcal{S}$, their corrupted version $\mathcal{S}^{m}$ with missing values, and the corresponding completed imputation results $\bar{\mathcal{S}}$., 
    The $Accuracy$ of samples dataset is computed with help with an \textit{oracle} model which is a highly accurate classification model trained on the \textit{real}, and complete, training dataset as we discussed earlier in Section ~\ref{sec:eval_accuracy}.
    
    the Semantic Repair (SR) of $\mathcal{S}^m$ is defined as:
    \begin{equation}
    \begin{split}
        \text{SR}(\mathcal{S}, \mathcal{S}^m, \bar{\mathcal{S}}) = \frac{Accuracy(\bar{\mathcal{S}}) - Accuracy(\mathcal{S}^m)}{Accuracy(\mathcal{S}) - Accuracy(\mathcal{S}^m)}
    \end{split}
    \end{equation}
    
\end{itemize}

Table ~\ref{tab:imputation_result}, compares the evaluation results of both MAE and \textit{semantic repair} metrics computed for different imputation methods on both the AFIb and HAR classification datasets under different scenarios of missing values. The result indicates that despite $\name$ has a higher MAE than other imputation methods, \name outperforms the other techniques in three out of four tests in the \textit{semantic repair} quality. This reflects the ability of \name to complete the signal in a semantically preserving way.

Finally, we would like to note that unlike other methods, \name was not intentionally trained to act as an imputer. The quality of \name results in data imputation is a byproduct of the quality of its data generation. The quality of \name in data imputation, in terms of both MAE and \texttt{Semantic Repair} can be further improved by training the \textit{encoder} and \textit{decoder} to reconstruct data samples from the noisy \textit{incomplete} inputs. i.e. using them as a `denoising autoencoder'~\cite{vincent2008extracting}.

  \section{Conclusion}
  \label{sec:discussion}
 In this paper, we presented \physgan a novel model architecture to train generative models for physiological sensor readings conditioned on their class labels. Compared to other baseline models which were trained using the vanilla maximum likelihood, GANs, and VAE training objectives, \physgan produces high-quality samples that attain good scores in both accuracy and diversity. We prove the utility of \physgan by showing its significant improvement over baseline algorithms in producing a synthetic dataset that we can use to train classification models with significantly higher test accuracies. Furthermore, we show that \physgan surpasses existing methods for sensor data imputation in filling the missing values with realistic values.
  Our future directions of research include improving the quality of \physgan to generate longer samples with higher sampling frequencies. Also, we will investigate how to train \physgan with differential privacy techniques to provide formal quantification of the limits of the identifiable and sensitive information from training examples disclosed by \physgan.

\section*{Acknowledgements}

The research reported in this paper was sponsored in part by: the CONIX Research Center, one of six centers in JUMP, a Semiconductor Research Corporation (SRC) program sponsored by DARPA; by the NIH mHealth Center for Discovery, Optimization and Translation of Temporally-Precise Interventions (mDOT) under award 1P41EB028242; by the National Science Foundation (NSF) under awards \# OAC-1640813 and CNS-1822935; and, by the IoBT REIGN Collaborative Research Alliance funded by the Army Research Laboratory (ARL) under Cooperative Agreement W911NF-17-2-0196. The views and conclusions contained in this document are those of the authors and should not be interpreted as representing the official policies, either expressed or implied, of the ARL, DARPA, NIH, NSF, SRC, or the U.S. Government. The U.S. Government is authorized to reproduce and distribute reprints for Government purposes notwithstanding any copyright notation here on.

\nocite{langley00}

\bibliography{main}

\begin{thebibliography}{69}
\providecommand{\natexlab}[1]{#1}
\providecommand{\url}[1]{\texttt{#1}}
\expandafter\ifx\csname urlstyle\endcsname\relax
  \providecommand{\doi}[1]{doi: #1}\else
  \providecommand{\doi}{doi: \begingroup \urlstyle{rm}\Url}\fi

\bibitem[Aghanavesi et~al.(2019)Aghanavesi, Bergquist, Nyholm, Senek, and
  Memedi]{aghanavesi2019motion}
Aghanavesi, S., Bergquist, F., Nyholm, D., Senek, M., and Memedi, M.
\newblock Motion sensor-based assessment of parkinson's disease motor symptoms
  during leg agility tests: results from levodopa challenge.
\newblock \emph{IEEE journal of biomedical and health informatics}, 2019.

\bibitem[Alzantot \& Srivastava(2019)Alzantot and Srivastava]{uclanesl_dp_wgan}
Alzantot, M. and Srivastava, M.
\newblock {Differential Privacy Synthetic Data Generation using WGANs}, 2019.

\bibitem[Alzantot et~al.(2017)Alzantot, Chakraborty, and
  Srivastava]{alzantot2017sensegen}
Alzantot, M., Chakraborty, S., and Srivastava, M.
\newblock Sensegen: A deep learning architecture for synthetic sensor data
  generation.
\newblock In \emph{2017 IEEE International Conference on Pervasive Computing
  and Communications Workshops (PerCom Workshops)}, pp.\  188--193. IEEE, 2017.

\bibitem[Anguita et~al.(2013)Anguita, Ghio, Oneto, Parra, and
  Reyes-Ortiz]{anguita2013public}
Anguita, D., Ghio, A., Oneto, L., Parra, X., and Reyes-Ortiz, J.~L.
\newblock A public domain dataset for human activity recognition using
  smartphones.
\newblock In \emph{Esann}, 2013.

\bibitem[Arjovsky et~al.(2017)Arjovsky, Chintala, and
  Bottou]{arjovsky2017wasserstein}
Arjovsky, M., Chintala, S., and Bottou, L.
\newblock Wasserstein gan.
\newblock \emph{arXiv preprint arXiv:1701.07875}, 2017.

\bibitem[Azur et~al.(2011)Azur, Stuart, Frangakis, and Leaf]{azur2011multiple}
Azur, M.~J., Stuart, E.~A., Frangakis, C., and Leaf, P.~J.
\newblock Multiple imputation by chained equations: what is it and how does it
  work?
\newblock \emph{International journal of methods in psychiatric research},
  20\penalty0 (1):\penalty0 40--49, 2011.

\bibitem[Bengio et~al.(2015)Bengio, Vinyals, Jaitly, and
  Shazeer]{bengio2015scheduled}
Bengio, S., Vinyals, O., Jaitly, N., and Shazeer, N.
\newblock Scheduled sampling for sequence prediction with recurrent neural
  networks.
\newblock In \emph{Advances in Neural Information Processing Systems}, pp.\
  1171--1179, 2015.

\bibitem[Bowman et~al.(2015)Bowman, Vilnis, Vinyals, Dai, Jozefowicz, and
  Bengio]{bowman2015generating}
Bowman, S.~R., Vilnis, L., Vinyals, O., Dai, A.~M., Jozefowicz, R., and Bengio,
  S.
\newblock Generating sentences from a continuous space.
\newblock \emph{arXiv preprint arXiv:1511.06349}, 2015.

\bibitem[Brock et~al.(2018)Brock, Donahue, and Simonyan]{brock2018large}
Brock, A., Donahue, J., and Simonyan, K.
\newblock Large scale gan training for high fidelity natural image synthesis.
\newblock \emph{arXiv preprint arXiv:1809.11096}, 2018.

\bibitem[Cao et~al.(2018)Cao, Wang, Li, Zhou, Li, and Li]{cao2018brits}
Cao, W., Wang, D., Li, J., Zhou, H., Li, L., and Li, Y.
\newblock Brits: bidirectional recurrent imputation for time series.
\newblock In \emph{Advances in Neural Information Processing Systems}, pp.\
  6775--6785, 2018.

\bibitem[Chen et~al.(2006)Chen, Kwong, Chang, Luk, and
  Bajcsy]{chen2006wearable}
Chen, J., Kwong, K., Chang, D., Luk, J., and Bajcsy, R.
\newblock Wearable sensors for reliable fall detection.
\newblock In \emph{2005 IEEE Engineering in Medicine and Biology 27th Annual
  Conference}, pp.\  3551--3554. IEEE, 2006.

\bibitem[Chen et~al.(1998)Chen, Beeferman, and Rosenfeld]{chen1998evaluation}
Chen, S.~F., Beeferman, D., and Rosenfeld, R.
\newblock Evaluation metrics for language models.
\newblock In \emph{DARPA Broadcast News Transcription and Understanding
  Workshop}, pp.\  275--280. Citeseer, 1998.

\bibitem[Chen et~al.(2016)Chen, Duan, Houthooft, Schulman, Sutskever, and
  Abbeel]{chen2016infogan}
Chen, X., Duan, Y., Houthooft, R., Schulman, J., Sutskever, I., and Abbeel, P.
\newblock Infogan: Interpretable representation learning by information
  maximizing generative adversarial nets.
\newblock In \emph{Advances in neural information processing systems}, pp.\
  2172--2180, 2016.

\bibitem[Cho et~al.(2014)Cho, Van~Merri{\"e}nboer, Bahdanau, and
  Bengio]{cho2014properties}
Cho, K., Van~Merri{\"e}nboer, B., Bahdanau, D., and Bengio, Y.
\newblock On the properties of neural machine translation: Encoder-decoder
  approaches.
\newblock \emph{arXiv preprint arXiv:1409.1259}, 2014.

\bibitem[Choi et~al.(2017)Choi, Biswal, Malin, Duke, Stewart, and
  Sun]{choi2017generating}
Choi, E., Biswal, S., Malin, B., Duke, J., Stewart, W.~F., and Sun, J.
\newblock Generating multi-label discrete patient records using generative
  adversarial networks.
\newblock \emph{arXiv preprint arXiv:1703.06490}, 2017.

\bibitem[Chung et~al.(2015)Chung, Gulcehre, Cho, and Bengio]{chung2015gated}
Chung, J., Gulcehre, C., Cho, K., and Bengio, Y.
\newblock Gated feedback recurrent neural networks.
\newblock In \emph{International Conference on Machine Learning}, pp.\
  2067--2075, 2015.

\bibitem[Demir \& Unal(2018)Demir and Unal]{demir2018patch}
Demir, U. and Unal, G.
\newblock Patch-based image inpainting with generative adversarial networks.
\newblock \emph{arXiv preprint arXiv:1803.07422}, 2018.

\bibitem[Dwork et~al.(2014)Dwork, Roth, et~al.]{dwork2014algorithmic}
Dwork, C., Roth, A., et~al.
\newblock The algorithmic foundations of differential privacy.
\newblock \emph{Foundations and Trends{\textregistered} in Theoretical Computer
  Science}, 9\penalty0 (3--4):\penalty0 211--407, 2014.

\bibitem[Esteban et~al.(2017)Esteban, Hyland, and R{\"a}tsch]{esteban2017real}
Esteban, C., Hyland, S.~L., and R{\"a}tsch, G.
\newblock Real-valued (medical) time series generation with recurrent
  conditional gans.
\newblock \emph{arXiv preprint arXiv:1706.02633}, 2017.

\bibitem[Fredrikson et~al.(2015)Fredrikson, Jha, and
  Ristenpart]{fredrikson2015model}
Fredrikson, M., Jha, S., and Ristenpart, T.
\newblock Model inversion attacks that exploit confidence information and basic
  countermeasures.
\newblock In \emph{Proceedings of the 22nd ACM SIGSAC Conference on Computer
  and Communications Security}, pp.\  1322--1333. ACM, 2015.

\bibitem[Friedman et~al.(2001)Friedman, Hastie, and
  Tibshirani]{friedman2001elements}
Friedman, J., Hastie, T., and Tibshirani, R.
\newblock \emph{The elements of statistical learning}, volume~1.
\newblock Springer series in statistics New York, 2001.

\bibitem[Goodfellow et~al.(2014)Goodfellow, Pouget-Abadie, Mirza, Xu,
  Warde-Farley, Ozair, Courville, and Bengio]{goodfellow2014generative}
Goodfellow, I., Pouget-Abadie, J., Mirza, M., Xu, B., Warde-Farley, D., Ozair,
  S., Courville, A., and Bengio, Y.
\newblock Generative adversarial nets.
\newblock In \emph{Advances in neural information processing systems}, pp.\
  2672--2680, 2014.

\bibitem[Graves(2013)]{graves2013generating}
Graves, A.
\newblock Generating sequences with recurrent neural networks.
\newblock \emph{arXiv preprint arXiv:1308.0850}, 2013.

\bibitem[Gulrajani et~al.(2017)Gulrajani, Ahmed, Arjovsky, Dumoulin, and
  Courville]{gulrajani2017improved}
Gulrajani, I., Ahmed, F., Arjovsky, M., Dumoulin, V., and Courville, A.~C.
\newblock Improved training of wasserstein gans.
\newblock In \emph{Advances in Neural Information Processing Systems}, pp.\
  5767--5777, 2017.

\bibitem[Ha \& Eck(2017)Ha and Eck]{ha2017neural}
Ha, D. and Eck, D.
\newblock A neural representation of sketch drawings.
\newblock \emph{arXiv preprint arXiv:1704.03477}, 2017.

\bibitem[Heusel et~al.(2017)Heusel, Ramsauer, Unterthiner, Nessler, and
  Hochreiter]{heusel2017gans}
Heusel, M., Ramsauer, H., Unterthiner, T., Nessler, B., and Hochreiter, S.
\newblock Gans trained by a two time-scale update rule converge to a local nash
  equilibrium.
\newblock In \emph{Advances in Neural Information Processing Systems}, pp.\
  6626--6637, 2017.

\bibitem[Higgins et~al.(2017)Higgins, Matthey, Pal, Burgess, Glorot, Botvinick,
  Mohamed, and Lerchner]{higgins2017beta}
Higgins, I., Matthey, L., Pal, A., Burgess, C., Glorot, X., Botvinick, M.,
  Mohamed, S., and Lerchner, A.
\newblock beta-vae: Learning basic visual concepts with a constrained
  variational framework.
\newblock \emph{ICLR}, 2\penalty0 (5):\penalty0 6, 2017.

\bibitem[Hu et~al.(2017)Hu, Yang, Liang, Salakhutdinov, and Xing]{hu2017toward}
Hu, Z., Yang, Z., Liang, X., Salakhutdinov, R., and Xing, E.~P.
\newblock Toward controlled generation of text.
\newblock In \emph{Proceedings of the 34th International Conference on Machine
  Learning-Volume 70}, pp.\  1587--1596. JMLR. org, 2017.

\bibitem[Husz{\'a}r(2015)]{huszar2015not}
Husz{\'a}r, F.
\newblock How (not) to train your generative model: Scheduled sampling,
  likelihood, adversary?gr.
\newblock \emph{arXiv preprint arXiv:1511.05101}, 2015.

\bibitem[Jordon et~al.(2018)Jordon, Yoon, and van~der Schaar]{jordon2018pate}
Jordon, J., Yoon, J., and van~der Schaar, M.
\newblock Pate-gan: generating synthetic data with differential privacy
  guarantees.
\newblock 2018.

\bibitem[Juefei-Xu et~al.(2018)Juefei-Xu, Dey, Boddeti, and
  Savvides]{juefei2018rankgan}
Juefei-Xu, F., Dey, R., Boddeti, V.~N., and Savvides, M.
\newblock Rankgan: A maximum margin ranking gan for generating faces.
\newblock In \emph{Asian Conference on Computer Vision}, pp.\  3--18. Springer,
  2018.

\bibitem[Karras et~al.(2019)Karras, Laine, and Aila]{karras2019style}
Karras, T., Laine, S., and Aila, T.
\newblock A style-based generator architecture for generative adversarial
  networks.
\newblock In \emph{Proceedings of the IEEE Conference on Computer Vision and
  Pattern Recognition}, pp.\  4401--4410, 2019.

\bibitem[Kingma \& Welling(2013)Kingma and Welling]{kingma2013auto}
Kingma, D.~P. and Welling, M.
\newblock Auto-encoding variational bayes.
\newblock \emph{arXiv preprint arXiv:1312.6114}, 2013.

\bibitem[Kingma et~al.()Kingma, Salimans, Jozefowicz, Chen, Sutskever, and
  Welling]{kingma1606improving}
Kingma, D.~P., Salimans, T., Jozefowicz, R., Chen, X., Sutskever, I., and
  Welling, M.
\newblock Improving variational inference with inverse autoregressive
  flow.(nips), 2016.
\newblock \emph{URL http://arxiv. org/abs/1606.04934}.

\bibitem[Ledig et~al.(2017)Ledig, Theis, Husz{\'a}r, Caballero, Cunningham,
  Acosta, Aitken, Tejani, Totz, Wang, et~al.]{ledig2017photo}
Ledig, C., Theis, L., Husz{\'a}r, F., Caballero, J., Cunningham, A., Acosta,
  A., Aitken, A., Tejani, A., Totz, J., Wang, Z., et~al.
\newblock Photo-realistic single image super-resolution using a generative
  adversarial network.
\newblock In \emph{Proceedings of the IEEE conference on computer vision and
  pattern recognition}, pp.\  4681--4690, 2017.

\bibitem[Lu et~al.(2018)Lu, Zhu, Zhang, Wang, and Yu]{lu2018neural}
Lu, S., Zhu, Y., Zhang, W., Wang, J., and Yu, Y.
\newblock Neural text generation: past, present and beyond.
\newblock \emph{arXiv preprint arXiv:1803.07133}, 2018.

\bibitem[Metz et~al.(2017)Metz, Poole, Pfau, and Sohl-Dickstein]{unrolledgan}
Metz, L., Poole, B., Pfau, D., and Sohl-Dickstein, J.
\newblock Unrolled generative adversarial networks.
\newblock 2017.
\newblock URL \url{https://openreview.net/pdf?id=BydrOIcle}.

\bibitem[Mikolov et~al.(2010)Mikolov, Karafi{\'a}t, Burget, {\v{C}}ernock{\`y},
  and Khudanpur]{mikolov2010recurrent}
Mikolov, T., Karafi{\'a}t, M., Burget, L., {\v{C}}ernock{\`y}, J., and
  Khudanpur, S.
\newblock Recurrent neural network based language model.
\newblock In \emph{Eleventh annual conference of the international speech
  communication association}, 2010.

\bibitem[Mirza \& Osindero(2014)Mirza and Osindero]{mirza2014conditional}
Mirza, M. and Osindero, S.
\newblock Conditional generative adversarial nets.
\newblock \emph{arXiv preprint arXiv:1411.1784}, 2014.

\bibitem[Oord et~al.(2016)Oord, Dieleman, Zen, Simonyan, Vinyals, Graves,
  Kalchbrenner, Senior, and Kavukcuoglu]{oord2016wavenet}
Oord, A. v.~d., Dieleman, S., Zen, H., Simonyan, K., Vinyals, O., Graves, A.,
  Kalchbrenner, N., Senior, A., and Kavukcuoglu, K.
\newblock Wavenet: A generative model for raw audio.
\newblock \emph{arXiv preprint arXiv:1609.03499}, 2016.

\bibitem[Papernot et~al.(2016)Papernot, Abadi, Erlingsson, Goodfellow, and
  Talwar]{papernot2016semi}
Papernot, N., Abadi, M., Erlingsson, U., Goodfellow, I., and Talwar, K.
\newblock Semi-supervised knowledge transfer for deep learning from private
  training data.
\newblock \emph{arXiv preprint arXiv:1610.05755}, 2016.

\bibitem[Park et~al.(2018)Park, Mohammadi, Gorde, Jajodia, Park, and
  Kim]{park2018data}
Park, N., Mohammadi, M., Gorde, K., Jajodia, S., Park, H., and Kim, Y.
\newblock Data synthesis based on generative adversarial networks.
\newblock \emph{Proceedings of the VLDB Endowment}, 11\penalty0 (10):\penalty0
  1071--1083, 2018.

\bibitem[Radford et~al.(2015)Radford, Metz, and
  Chintala]{radford2015unsupervised}
Radford, A., Metz, L., and Chintala, S.
\newblock Unsupervised representation learning with deep convolutional
  generative adversarial networks.
\newblock \emph{arXiv preprint arXiv:1511.06434}, 2015.

\bibitem[Ravuri \& Vinyals(2019)Ravuri and Vinyals]{ravuri2019classification}
Ravuri, S. and Vinyals, O.
\newblock Classification accuracy score for conditional generative models.
\newblock \emph{arXiv preprint arXiv:1905.10887}, 2019.

\bibitem[Razavi et~al.(2019{\natexlab{a}})Razavi, Oord, Poole, and
  Vinyals]{razavi2019preventing}
Razavi, A., Oord, A. v.~d., Poole, B., and Vinyals, O.
\newblock Preventing posterior collapse with delta-vaes.
\newblock \emph{arXiv preprint arXiv:1901.03416}, 2019{\natexlab{a}}.

\bibitem[Razavi et~al.(2019{\natexlab{b}})Razavi, Oord, and
  Vinyals]{razavi2019generating}
Razavi, A., Oord, A. v.~d., and Vinyals, O.
\newblock Generating diverse high-fidelity images with vq-vae-2.
\newblock \emph{arXiv preprint arXiv:1906.00446}, 2019{\natexlab{b}}.

\bibitem[Reed et~al.(2016)Reed, Akata, Yan, Logeswaran, Schiele, and
  Lee]{reed2016generative}
Reed, S., Akata, Z., Yan, X., Logeswaran, L., Schiele, B., and Lee, H.
\newblock Generative adversarial text to image synthesis.
\newblock \emph{arXiv preprint arXiv:1605.05396}, 2016.

\bibitem[Roberts et~al.(2018)Roberts, Engel, Raffel, Hawthorne, and
  Eck]{roberts2018hierarchical}
Roberts, A., Engel, J., Raffel, C., Hawthorne, C., and Eck, D.
\newblock A hierarchical latent vector model for learning long-term structure
  in music.
\newblock \emph{arXiv preprint arXiv:1803.05428}, 2018.

\bibitem[Salimans et~al.(2016)Salimans, Goodfellow, Zaremba, Cheung, Radford,
  and Chen]{salimans2016improved}
Salimans, T., Goodfellow, I., Zaremba, W., Cheung, V., Radford, A., and Chen,
  X.
\newblock Improved techniques for training gans.
\newblock In \emph{Advances in neural information processing systems}, pp.\
  2234--2242, 2016.

\bibitem[Salvador \& Chan(2007)Salvador and Chan]{salvador2007toward}
Salvador, S. and Chan, P.
\newblock Toward accurate dynamic time warping in linear time and space.
\newblock \emph{Intelligent Data Analysis}, 11\penalty0 (5):\penalty0 561--580,
  2007.

\bibitem[Shokri et~al.(2017)Shokri, Stronati, Song, and
  Shmatikov]{shokri2017membership}
Shokri, R., Stronati, M., Song, C., and Shmatikov, V.
\newblock Membership inference attacks against machine learning models.
\newblock In \emph{2017 IEEE Symposium on Security and Privacy (SP)}, pp.\
  3--18. IEEE, 2017.

\bibitem[Silberman \& Kahn(2011)Silberman and Kahn]{silberman2011burdens}
Silberman, G. and Kahn, K.~L.
\newblock Burdens on research imposed by institutional review boards: the state
  of the evidence and its implications for regulatory reform.
\newblock \emph{The Milbank Quarterly}, 89\penalty0 (4):\penalty0 599--627,
  2011.

\bibitem[Smith \& Smith(2020)Smith and Smith]{smith2020conditional}
Smith, K.~E. and Smith, A.~O.
\newblock Conditional gan for timeseries generation.
\newblock \emph{arXiv preprint arXiv:2006.16477}, 2020.

\bibitem[Srivastava et~al.(2017)Srivastava, Valkov, Russell, Gutmann, and
  Sutton]{srivastava2017veegan}
Srivastava, A., Valkov, L., Russell, C., Gutmann, M.~U., and Sutton, C.
\newblock Veegan: Reducing mode collapse in gans using implicit variational
  learning.
\newblock In \emph{Advances in Neural Information Processing Systems}, pp.\
  3308--3318, 2017.

\bibitem[Sutton et~al.(2000)Sutton, McAllester, Singh, and
  Mansour]{sutton2000policy}
Sutton, R.~S., McAllester, D.~A., Singh, S.~P., and Mansour, Y.
\newblock Policy gradient methods for reinforcement learning with function
  approximation.
\newblock In \emph{Advances in neural information processing systems}, pp.\
  1057--1063, 2000.

\bibitem[Theis et~al.(2015)Theis, Oord, and Bethge]{theis2015note}
Theis, L., Oord, A. v.~d., and Bethge, M.
\newblock A note on the evaluation of generative models.
\newblock \emph{arXiv preprint arXiv:1511.01844}, 2015.

\bibitem[Vincent et~al.(2008)Vincent, Larochelle, Bengio, and
  Manzagol]{vincent2008extracting}
Vincent, P., Larochelle, H., Bengio, Y., and Manzagol, P.-A.
\newblock Extracting and composing robust features with denoising autoencoders.
\newblock In \emph{Proceedings of the 25th international conference on Machine
  learning}, pp.\  1096--1103. ACM, 2008.

\bibitem[Wang et~al.(2018)Wang, Chen, Gu, Xiao, and Pan]{wang2018sensorygans}
Wang, J., Chen, Y., Gu, Y., Xiao, Y., and Pan, H.
\newblock Sensorygans: An effective generative adversarial framework for
  sensor-based human activity recognition.
\newblock In \emph{2018 International Joint Conference on Neural Networks
  (IJCNN)}, pp.\  1--8. IEEE, 2018.

\bibitem[Wang \& Wan(2018)Wang and Wan]{wang2018sentigan}
Wang, K. and Wan, X.
\newblock Sentigan: Generating sentimental texts via mixture adversarial
  networks.
\newblock In \emph{IJCAI}, pp.\  4446--4452, 2018.

\bibitem[with the special contribution of~the European Heart Rhythm
  Association~(EHRA) et~al.(2010)with the special contribution of~the European
  Heart Rhythm Association~(EHRA), by~the European Association~for
  Cardio-Thoracic Surgery~(EACTS), Members, Camm, Kirchhof, Lip, Schotten,
  Savelieva, Ernst, Van~Gelder, et~al.]{developed2010guidelines}
with the special contribution of~the European Heart Rhythm Association~(EHRA),
  D., by~the European Association~for Cardio-Thoracic Surgery~(EACTS), E.,
  Members, A.~F., Camm, A.~J., Kirchhof, P., Lip, G.~Y., Schotten, U.,
  Savelieva, I., Ernst, S., Van~Gelder, I.~C., et~al.
\newblock Guidelines for the management of atrial fibrillation: the task force
  for the management of atrial fibrillation of the european society of
  cardiology (esc).
\newblock \emph{European heart journal}, 31\penalty0 (19):\penalty0 2369--2429,
  2010.

\bibitem[Xu \& Veeramachaneni(2018)Xu and Veeramachaneni]{xu2018synthesizing}
Xu, L. and Veeramachaneni, K.
\newblock Synthesizing tabular data using generative adversarial networks.
\newblock \emph{arXiv preprint arXiv:1811.11264}, 2018.

\bibitem[Xu et~al.(2018)Xu, Zhang, Huang, Zhang, Gan, Huang, and
  He]{xu2018attngan}
Xu, T., Zhang, P., Huang, Q., Zhang, H., Gan, Z., Huang, X., and He, X.
\newblock Attngan: Fine-grained text to image generation with attentional
  generative adversarial networks.
\newblock In \emph{Proceedings of the IEEE Conference on Computer Vision and
  Pattern Recognition}, pp.\  1316--1324, 2018.

\bibitem[Yahi et~al.(2017)Yahi, Vanguri, Elhadad, and
  Tatonetti]{yahi2017generative}
Yahi, A., Vanguri, R., Elhadad, N., and Tatonetti, N.~P.
\newblock Generative adversarial networks for electronic health records: A
  framework for exploring and evaluating methods for predicting drug-induced
  laboratory test trajectories.
\newblock \emph{arXiv preprint arXiv:1712.00164}, 2017.

\bibitem[Y{\i}ld{\i}r{\i}m et~al.(2018)Y{\i}ld{\i}r{\i}m, P{\l}awiak, Tan, and
  Acharya]{yildirim2018arrhythmia}
Y{\i}ld{\i}r{\i}m, {\"O}., P{\l}awiak, P., Tan, R.-S., and Acharya, U.~R.
\newblock Arrhythmia detection using deep convolutional neural network with
  long duration ecg signals.
\newblock \emph{Computers in biology and medicine}, 102:\penalty0 411--420,
  2018.

\bibitem[Yu et~al.(2017)Yu, Zhang, Wang, and Yu]{yu2017seqgan}
Yu, L., Zhang, W., Wang, J., and Yu, Y.
\newblock Seqgan: Sequence generative adversarial nets with policy gradient.
\newblock In \emph{Thirty-First AAAI Conference on Artificial Intelligence},
  2017.

\bibitem[Zhang et~al.(2018)Zhang, Goodfellow, Metaxas, and
  Odena]{zhang2018self}
Zhang, H., Goodfellow, I., Metaxas, D., and Odena, A.
\newblock Self-attention generative adversarial networks.
\newblock \emph{arXiv preprint arXiv:1805.08318}, 2018.

\bibitem[Zhang \& Alshurafa(2020)Zhang and Alshurafa]{zhang2020deep}
Zhang, S. and Alshurafa, N.
\newblock Deep generative cross-modal on-body accelerometer data synthesis from
  videos.
\newblock In \emph{Adjunct Proceedings of the 2020 ACM International Joint
  Conference on Pervasive and Ubiquitous Computing and Proceedings of the 2020
  ACM International Symposium on Wearable Computers}, pp.\  223--227, 2020.

\bibitem[Zhang et~al.(2016)Zhang, Gan, and Carin]{zhang2016generating}
Zhang, Y., Gan, Z., and Carin, L.
\newblock Generating text via adversarial training.
\newblock In \emph{NIPS workshop on Adversarial Training}, volume~21, 2016.

\bibitem[Zhu et~al.(2018)Zhu, Lu, Zheng, Guo, Zhang, Wang, and
  Yu]{zhu2018texygen}
Zhu, Y., Lu, S., Zheng, L., Guo, J., Zhang, W., Wang, J., and Yu, Y.
\newblock Texygen: A benchmarking platform for text generation models.
\newblock In \emph{The 41st International ACM SIGIR Conference on Research \&
  Development in Information Retrieval}, pp.\  1097--1100. ACM, 2018.

\end{thebibliography}
\bibliographystyle{icml2021}

\appendix
\section{Visual Comparison of Synthetic Samples From Different Generative Models}
\begin{figure*}[!hbp]
    \centering
    \includegraphics[scale=0.3]{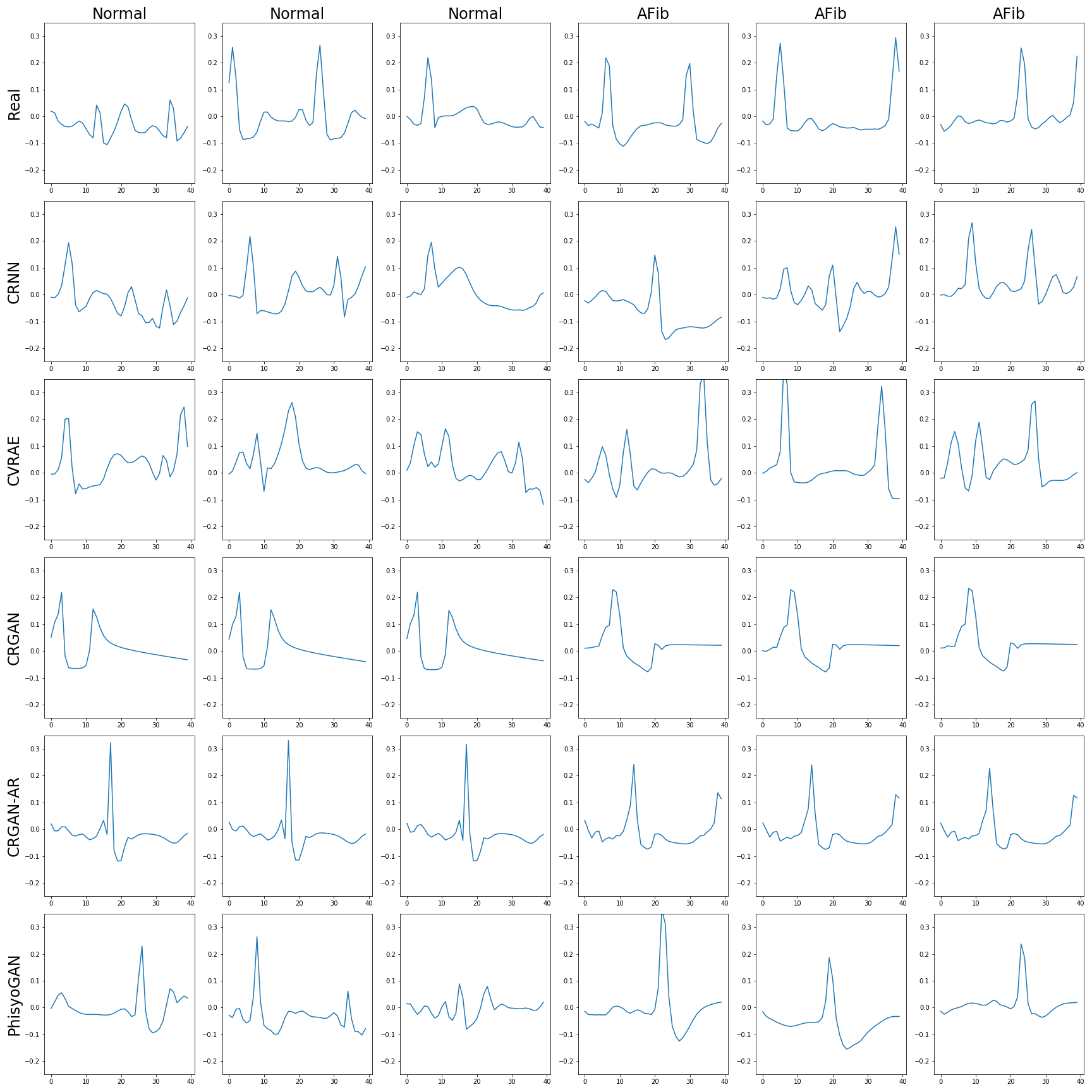}
\caption{Random samples of the real data (top row) and synthetic data generated by different generative models on the AFib dataset. The title of each column indicate the class label of the samples. Samples generated with length=40 time steps.}
    \label{fig:ecg_all_short}
\end{figure*}

\begin{figure*}[!hbp]
    \centering
    \includegraphics[scale=0.3]{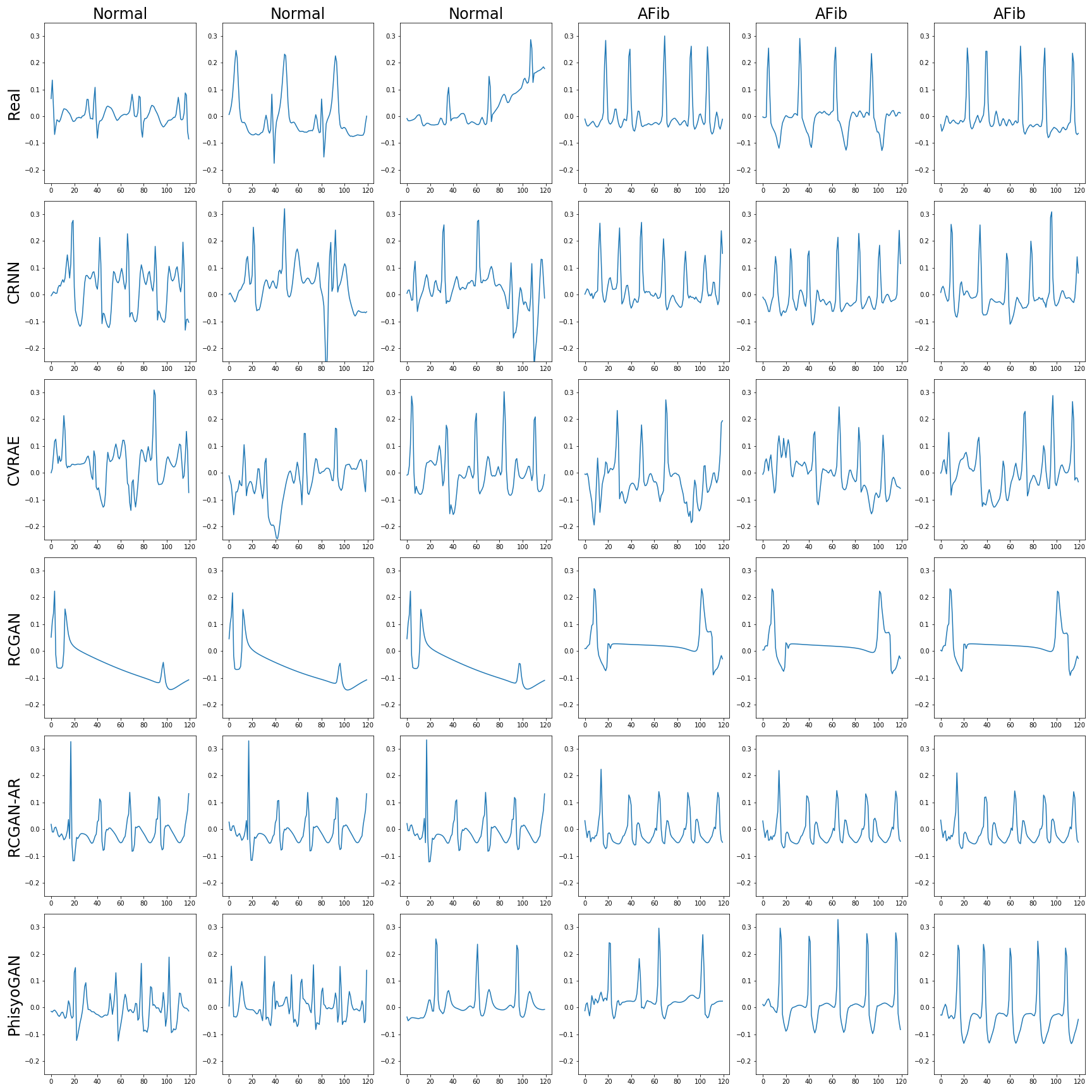}
\caption{Random samples of the real data (top row) and synthetic data generated by different generative models on the AFib dataset. The title of each column indicate the class label of the samples. Samples generated with length=120 time steps to compare the effectiveness of different models to generate long sequences.}
    \label{fig:ecg_all_long}
\end{figure*}

\begin{figure*}[!hbp]
    \centering
    \includegraphics[scale=0.3]{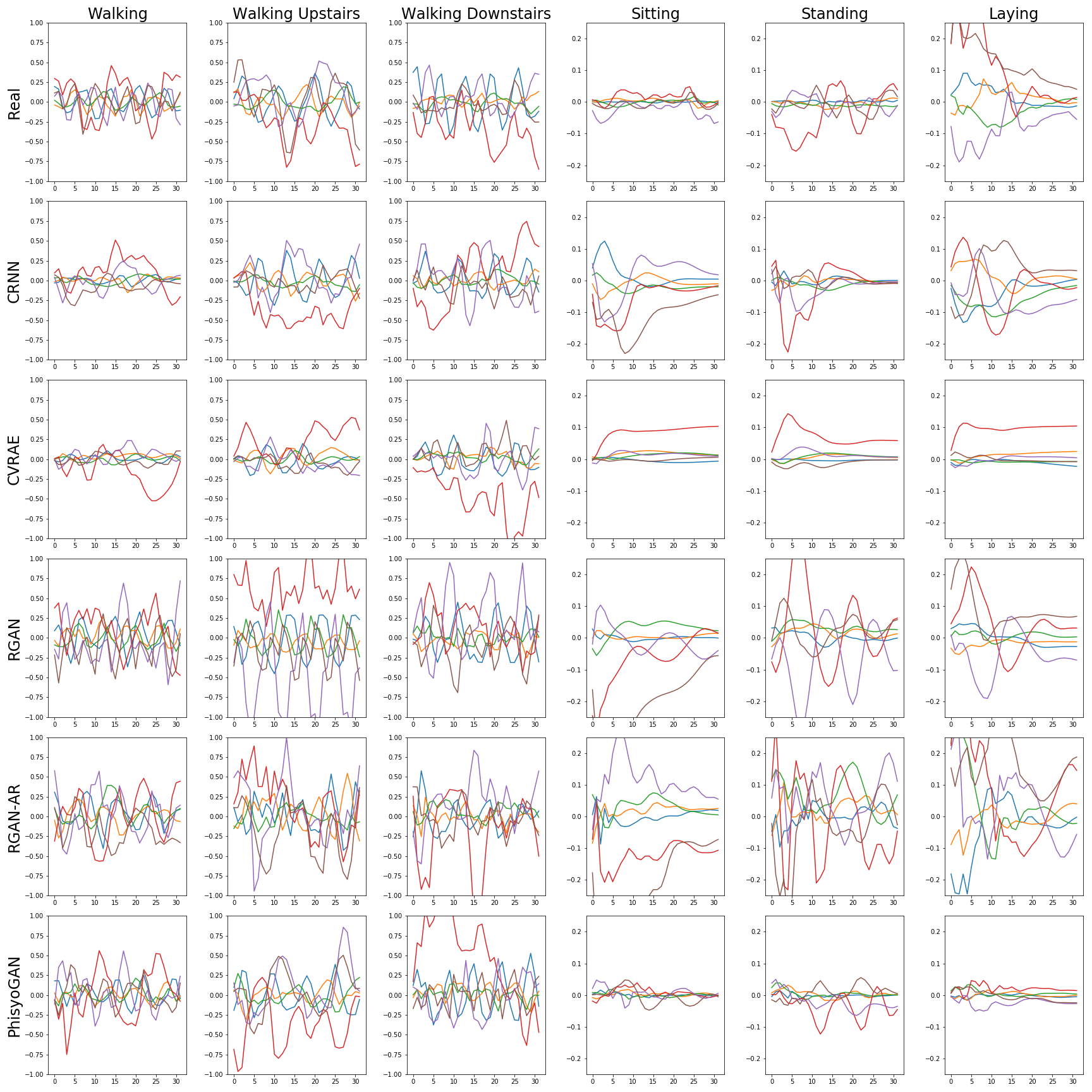}
\caption{Random samples of the real data (top row) and synthetic data generated by different generative models on the HAR dataset. The title of each column indicate the class label of the samples.}
    \label{fig:har_all}
\end{figure*}

Figures ~\ref{fig:ecg_all_short} and ~\ref{fig:ecg_all_long} provides a visual comparison between samples taken from the real dataset (top row) and samples generated by different generative models. The length of generated samples was 40 and 120 in Figure ~\ref{fig:ecg_all_short} and ~\ref{fig:ecg_all_long}, respectively. Figure ~\ref{fig:har_all} shows a comparison between the real data and synthetic data generated by different generative models on the HAR dataset.

\end{document}